\documentclass{biometrika}

\usepackage{amsmath}

\usepackage{times}
\usepackage{bm}
\usepackage{natbib}

\usepackage[plain,noend]{algorithm2e}

\makeatletter
\renewcommand{\algocf@captiontext}[2]{#1\algocf@typo. \AlCapFnt{}#2} % text of caption
\renewcommand{\AlTitleFnt}[1]{#1\unskip}% default definition
\def\@algocf@capt@plain{top}
\renewcommand{\algocf@makecaption}[2]{%
  \addtolength{\hsize}{\algomargin}%
  \sbox\@tempboxa{\algocf@captiontext{#1}{#2}}%
  \ifdim\wd\@tempboxa >\hsize%     % if caption is longer than a line
    \hskip .5\algomargin%
    \parbox[t]{\hsize}{\algocf@captiontext{#1}{#2}}% then caption is not centered
  \else%
    \global\@minipagefalse%
    \hbox to\hsize{\box\@tempboxa}% else caption is centered
  \fi%
  \addtolength{\hsize}{-\algomargin}%
}
\makeatother

\def\T{{ \mathrm{\scriptscriptstyle T} }}
\def\v{{\varepsilon}}

\begin{document}

\jname{Biometrika}
%% The year, volume, and number are determined on publication
%\jyear{}
%\jvol{}
%\jnum{}
%% The \doi{...} and \accessdate commands are used by the production team
%\doi{10.1093/biomet/asm023}
\accessdate{}
\copyrightinfo{\Copyright\ 2014 Biometrika Trust\goodbreak {\em Printed in Great Britain}}

%% These dates are usually set by the production team
\received{August 2014}
%\revised{September 2012}

%% The left and right page headers are defined here:
\markboth{T. Kunihama \and D. B. Dunson}{Biometrika style}

%% Here are the title, author names and addresses
\title{Nonparametric Bayes inference on conditional independence}

\author{T. Kunihama \and D. B. Dunson}
\affil{Department of Statistical Science, Duke University, Durham, North Carolina 27708-0251, U.S.A. \email{tsuyoshi.kunihama@duke.edu} \email{dunson@duke.edu}}

\maketitle

\begin{abstract}
In many application areas, a primary focus is on assessing evidence in the data refuting the assumption of independence of $Y$ and $X$ conditionally on $Z$, with $Y$ response variables, 
$X$ predictors of interest, and $Z$ covariates.  Ideally, one would have methods available that avoid parametric assumptions, allow $Y, X, Z$ to be random variables on arbitrary spaces with arbitrary dimension, and accommodate rapid consideration of different candidate predictors.  As a formal decision-theoretic approach has clear disadvantages in this context, we instead rely on an encompassing nonparametric Bayes model for the joint distribution of $Y$, $X$ and $Z$, with conditional mutual information used as a summary of the strength of conditional dependence.  The implementation relies on a single Markov chain Monte Carlo run under the encompassing model, with conditional mutual informations for candidate models calculated as a byproduct.  We provide asymptotic theory supporting the approach, and apply the method to variable selection.  The methods are illustrated through simulations and criminology applications.
\end{abstract}


\section{Introduction}

One of the canonical problems in statistics is to assess whether or not $Y$ is conditionally independent of $X$ given $Z$, expressed as $Y \perp X \mid Z$.  In general, $Y \in \mathcal{Y}$ is a response, $X \in \mathcal{X}$ are predictors of interest, $Z \in \mathcal{Z}$ are adjustment variables or covariates, and the variables can be multivariate and have a variety of measurement scales and domains.  There is a rich literature on testing of conditional independence in parametric models; often this corresponds to testing whether a vector of regression coefficients for the $X$ variables are equal to zero.  However, much less consideration has been given to this problem from a nonparametric perspective, particularly from a model-based Bayesian perspective.

In the frequentist literature, various nonparametric methods of testing conditional independence have been proposed, relying on different expressions of conditional independence with characteristic functions \citep{SuWhite07}, probability density functions \citep{SuWhite08, Perez08},  distribution functions \citep{SethPrincipe10, GyorfiWalk12}, copula densities \citep{BouezmarniRomboutsTaamouti12} and kernel methods \citep{FukumizuGrettonSun08}.  \cite{SethPrincipe12} develop an asymmetric measure of conditional independence based on cumulative distribution functions.  Also, \cite{Song09} constructs a test using Rosenblatt-transforms of random variables.  However, these approaches do not work well in the case where the dimension of data is not small and the performance can be heavily affected by the choice of free parameters \citep{SethPrincipe12b}.

A rich variety of Bayesian nonparametric models have been proposed for joint and conditional distributions, ranging from Dirichlet process mixtures \citep{Lo84, WestMullerEscobar94, EscobarWest95, MullerErkanliWest96} to kernel stick-breaking processes \citep{DunsonPark08, AnWangShterevWangCarinDunson08}.  
However, such models do not allow testing of conditional independence relationships.  A Bayesian decision-theoretic approach to the problem would (i) define a list of possible conditional independence relationships {\em a priori}, (ii) specify a nonparametric Bayes model for each relationship, (iii) calculate marginal likelihoods, and (iv) choose the relationship having minimal expected loss.  However, a number of major practical problems arise.  It is in general not straightforward to define a nonparametric Bayes model, which has full support on the space of distributions satisfying a particular conditional independence relationship, making (ii) problematic.  Even if one could define appropriate models, (iii) is an issue due to the intractability of accurately approximating marginal likelihoods in infinite-dimensional Bayesian models.  Also, even if (ii)-(iii) could be achieved, the behavior of marginal likelihoods in infinite-dimensional models is poorly understood, and misleading results are possible as mentioned in a 2012 Ohio State University PhD thesis by L. Pingbo.

There is a small literature on Bayesian nonparametric methods for variable selection \citep{ChungDunson09, Ma13, ReichKalendraStorlie12}, attempting to follow the above strategy in specialized settings.  However, there has been essentially no theoretic justification for these methods, and the practical implementation is limited to low-dimensional settings.  In this article, we propose a substantially different approach.  In particular, instead of attempting to select between different {\em exact} conditional independence relationships, we define an encompassing Bayesian nonparametric model, which is sufficiently flexible to approximate any relationship.  We then use conditional mutual information as a scalar summary of the strength of departure from a particular conditional independence relationship.  We estimate the conditional mutual information relying on a functional of the encompassing model and the empirical measure.  The proposed framework is useful for rapid screening of variables that add significantly to prediction, and can be implemented easily leveraging on Markov chain Monte Carlo algorithms for the encompassing model.  Based on empirical process theory, we show that the proposed method consistently selects conditionally dependent predictors under appropriate conditions. 

%%%%%%%%%%%%%%%%%%%%%%%%%%%%%%%%%%%%%%%%%%%%
\section{Inference on conditional independence}
%%%%%%%%%%%%%%%%%%%%%%%%%%%%%%%%%%%%%%%%%%%%

%%%%%%%%%%%%%%%%%%%%%%%%%%%%%%%%%%%%%%%%%%%%%%%%%
\subsection{Conditional mutual information}

Let $Y$, $X$ and $Z$ be univariate or multivariate random variables where each element can have any type of scale and domain.  We also let $f(y,x,z)$ denote the joint density of $Y$, $X$ and $Z$ with respect to a product measure $\xi$.  The marginal densities we use below are denoted by $f(y,z)$, $f(x,z)$ and $f(z)$.  Suppose the primary interest is in assessing if $Y$ and $X$ are conditionally independent given $Z$.  Relying on the joint density, $Y \perp X \mid Z$ can be equivalently expressed as
\begin{align*}
f(y,x,z) f(z) = f(y,z) f(x,z),  
%\label{eq:cond-ind}
\end{align*}
for all $(y,x,z)$ in the support of $f$.

In information theory, conditional mutual information measures the strength of functional relationship between $Y$ and $X$ given $Z$ \citep{Wyner78, Joe89, MacKay03, CoverThomas06},
 
\begin{align*}
\zeta = \int f(y,x,z) \log \frac{f(y,x,z) f(z)}{f(y,z) f(x,z)} d \xi.
\end{align*}
Letting $KL(p,q)=\int p \log(p/q)$ denote the Kullback-Leibler divergence, $\zeta = KL\{ f(y,x,z), f(y,z) f(x,z)/f(z)\}$, which is always non-negative.  In general, $\zeta = 0$ if and only if $Y \perp X \mid Z$, while large values of $\zeta$ indicate substantial violations of conditional independence with an approximate functional relationship between $Y$ and $X$ given $Z$.

\subsection{Empirical Bayes estimation of conditional mutual information} \label{proposed-test}

Let $P_0$ denote a {\em true} data-generating probability having density $f_0 \in L_{\xi}$, with $L_{\xi}$ the set of all probability densities with respect to a measure $\xi$.  Let $\Pi$ denote a prior probability on $L_{\xi}$ with $\Pi(\mathcal{F})=1$ for $\mathcal{F} \subset L_{\xi}$.  
%A countable product of $P_0$ is denoted by $P_0^{\infty}$.  
Data $D_n$ consist of independently identically distributed observations $(y_i,x_i,z_i)$ from $P_0$ with $i=1,\ldots,n$.  Let $\zeta_0$ be the conditional mutual information induced by the true data-generating distribution,
\begin{align*}
\zeta_0 = \int \log \frac{f_0(y,x,z) f_0(z)}{f_0(y,z) f_0(x,z)} dP_0 =  \int f_0(y,x,z) \log \frac{f_0(y,x,z) f_0(z)}{f_0(y,z) f_0(x,z)} d \xi.
%\label{eq:cmi-true}
\end{align*}
As noted above, $Y \perp X \,|\, Z$ if and only if $\zeta_0=0$.  To estimate $\zeta_0$, we rely on an encompassing nonparametric Bayes model for the joint density $f_0$.  First, we define a function $\zeta(\cdot,\cdot)$ of a joint density $f \in L_{\xi}$ and a probability measure $P$ on $\mathcal{X}\times \mathcal{Y}\times \mathcal{Z}$ as
\begin{align}
\zeta(f,P) = \int \log \frac{f(y,x,z) f(z)}{f(y,z)f(x,z)} dP.
\label{eq:function}
\end{align} 
Using this function,  $\zeta_0$ can be expressed as $\zeta(f_0,P_0)$.  Intuitively, if $f$ and $P$ are close to $f_0$ and $P_0$ in some sense, $\zeta(f,P)$ can approximate $\zeta_0$ well.  In general, a probability measure $P$ having a density leads to a computationally intractable $\zeta(f,P)$ because of the difficulty in evaluating its integral.  Therefore, we utilize the empirical measure as an estimate of $P_0$,
\begin{align*}
P_n = \frac{1}{n} \sum_{i=1}^n \delta_{(y_i,x_i,z_i)},
\end{align*}
where $\delta_{(y,x,z)}$ is the Dirac measure concentrated at $(y,x,z)$.  The empirical measure $P_n$ is a consistent estimate of $P_0$ in that $P_n(A)\rightarrow P_0(A)$ almost surely for any $A$ by the strong law of large numbers.  Then, we let 
\begin{align}
\zeta(f, P_n) = \int \log \frac{f(y,x,z) f(z)}{f(y,z)f(x,z)} dP_n  = \frac{1}{n} \sum_{i=1}^n \log \frac{f(y_i,x_i,z_i) f(z_i)}{f(y_i,z_i)f(x_i,z_i)}, \ \ f\in \mathcal{F},
\label{eq:estimate}
\end{align}
where $\zeta(f,P_n) \in \Re$ and, for any fixed $f \in \mathcal{F}$, $\zeta(f,P_n) \to \zeta(f,P_0)$ almost surely $P_0^{\infty}$ by the law of large numbers.  By using the empirical measure $P_n$ for $P_0$ while defining a nonparametric Bayes encompassing prior for the joint density $f$, we define an empirical Bayes approach that induces a posterior on $\zeta$ accounting for uncertainty.  In finite samples this posterior assigns non-zero probability to $\zeta < 0$, which results because $P_n$ does not exactly correspond to the measure induced from the density $f$.

Plugging in the empirical measure $P_n$, expression (\ref{eq:estimate}) for the conditional mutual information depends on the unknown joint density $f$ and corresponding marginals.  Updating prior $f \sim \Pi$ with data $(y_i,x_i,z_i), i=1,\ldots,n$, we obtain a posterior quantifying our current state of knowledge about the density $f$.  We can obtain samples from this posterior by running Markov chain Monte Carlo for the encompassing model ignoring any conditional independence structure.  Then, to marginalize $f$ out of expression (\ref{eq:estimate}) and obtain an empirical Bayes estimate of $\zeta_0$, we simply use Monte Carlo integration.  In particular, for each draw from the posterior, we compute and save $\zeta(f,P_n)$.  The resulting draws of $\zeta$ are from the induced empirical Bayes posterior of the conditional mutual information; we use this posterior as the basis for our inferences.  

Under our asymptotic theory  below, as $n$ increases the posterior of $\zeta(f,P_n)$ will be increasingly concentrated around the true conditional mutual information $\zeta_0$.  Therefore, if $\zeta_0$ is not close to zero, zero should locate in the left tail of the distribution of $\zeta(f,P_n)$.  We consider the posterior probability of $\zeta(f,P_n)$ being positive as a weight of evidence of violations of conditional independence.  The posterior probability can be estimated by $(1/R)\sum_{r=1}^R 1\{\zeta(f^{(r)},P_n) > 0\}$ where $R$ is the number of Markov chain Monte Carlo iterations after the burn-in period, $1\{\cdot\}$ is an indicator function and $f^{(r)}$ is the joint density under the encompassing model at the $r$th iteration.

\subsection{Theoretic support}

The next theorem provides sufficient conditions under which the posterior of $\zeta(f, P_n)$ concentrates on arbitrarily small neighborhoods of $\zeta_0$ as the sample size increases.  
\begin{theorem}
\label{theorem1}
Suppose for any $\epsilon>0$,
\begin{align}
\Pi \left[ KL\{f_0(y,x,z),f(y,x,z)\} < \epsilon  \right] > 0
\label{eq:kl-support}
\end{align}
and the following classes of functions
\begin{align*}
 \left\{ \log \frac{f_0(y,x,z)}{f(y,x,z)}, f\in \mathcal{F}\right\},   \left\{ \log \frac{f_0(y,z)}{f(y,z)}, f\in \mathcal{F} \right\}, \left\{ \log \frac{f_0(x,z)}{f(x,z)}, f\in \mathcal{F} \right\}, \left\{ \log \frac{f_0(z)}{f(z)}, f\in \mathcal{F} \right\},
\end{align*}
are $P_0$-Glivenko-Cantelli. 
Then, for any $\epsilon'>0$
\begin{align*}
\Pi \left\{ |\zeta(f,P_n) - \zeta_0| < \epsilon' \mid D_n \right\} \rightarrow 1, \ \ \text{{\it almost surely $P_0^{\infty}$}}.
%\label{eq:threshold}
\end{align*}
\end{theorem}
The proof is in the Appendix.  The condition (\ref{eq:kl-support}) means the true data-generating density is in the Kullback-Leibler support of the prior.  Such support conditions are standard for Bayesian nonparametric models, and are routinely employed in theorems of posterior asymptotics \citep{GhosalGhoshRamamoorthi99, GhoshRamamoorthi03, Tokdar06}.  \cite{WuGhosal08} discuss the Kullback-Leibler property for various types of kernels in Dirichlet process mixture models.  As for the Glivenko-Cantelli class, theoretical properties of the class have been studied in empirical process theory \citep{vanderVaartWellner96, Kosorok08}.  It is a wide class of functions such that the law of large numbers holds uniformly over the space.

%%%%%%%%%%%%%%%%%%%%%%%%%%%%%%%%%%%%%%%%%%%%%%%%%%%%%%%%%%%%%%%%%%%%%%%%%%%%
\subsection{Variable selection} \label{variable selection}
%%%%%%%%%%%%%%%%%%%%%%%%%%%%%%%%%%%%%%%%%%%%%%%%%%%%%%%%%%%%%%%%%%%%%%%%%%%%

Suppose we have a univariate response $Y\in \mathcal{Y}$ and vector of predictors $X = (X_1,\ldots,X_p)^{\T}$.  Conditional mutual information provides a measure of how much information a particular predictor $X_j$ adds when included in a model already containing the predictors in $X_{-j} = (X_1,\ldots,X_{j-1},X_{j+1},\ldots,X_{p})^{\T}$.  We can potentially use our method for predictive variable selection, conducting a search for the smallest subset of variables $\gamma \subset \{1,\ldots,p\}$ such that there is no evidence of departure from 
$Y \perp X_{-\gamma} \mid X_{\gamma}$, with $X_{\gamma} = \{ X_j: j \in \gamma \}$ and $X_{-\gamma} = \{ X_j: j \not\in \gamma\}$.  However, instead of identifying parsimonious models for predicting $Y$, 
we focus here on selecting predictors that add significantly to models containing all other predictors.  This reduces the search from $2^p$ to $p$, while still producing results of inferential interest.  The computational savings come at the potential expense of excluding a set of important predictors containing redundant information about $Y$.
 
Let $\zeta_{0,j}$ be the true conditional mutual information for $Y\perp X_j\mid X_{-j}$. Let $\zeta_j(f,P_n)$ denote the value of $\zeta(f,P_n)$ in expression (2) with $x$ the $j$th predictor and $z$ the other predictors.  Posterior computation proceeds as in subsection \ref{proposed-test}.  We use the posterior probability of $\zeta_j(f,P_n)>0$ as evidence of violating $Y \perp X_j \mid X_{-j}$ for $j=1,\ldots,p$, selecting predictors having large probabilities.  This method is justified by the next theorem, which indicates zero should be in the left tail of the posterior distribution of $\zeta_j(f,P_n)$ under conditional dependence.

We show posterior consistency of $\zeta_j(f, P_n)$ to $\zeta_{0,j}$ under appropriate conditions.  Theorem 2 modifies Theorem 1 to the case  of measuring dependence between each predictor and the response, adjusting for all other predictors as covariates.  The difference from Theorem 1 is the Glivenko-Cantelli class condition depends on $j$.  Also, Theorem 2 states the posterior of $\zeta_j(f,P_n)$ will concentrate on $\zeta_{0,j}$ uniformly over $j$ as the sample size increases, allowing us to avoid multiple separate pairwise comparisons.  The proof is similar to that of Theorem 1 and given in the Supplementary Material.  
\begin{theorem}
\label{theorem2}
Suppose for any $\epsilon>0$,
\begin{align}
\Pi \left[ KL\{f_0(y,x),f(y,x)\} < \epsilon  \right] > 0
\label{eq:kl-support2}
\end{align}
and the following classes of functions
\begin{align*}
 \left\{ \log \frac{f_0(y,x)}{f(y,x)},  f\in \mathcal{F}\right\},   \left\{ \log \frac{f_0(x)}{f(x)}, f\in \mathcal{F} \right\},
 \left\{ \log \frac{f_0(y,x_{-j})}{f(y,x_{-j})}, f\in \mathcal{F} \right\},  \left\{ \log \frac{f_0(x_{-j})}{f(x_{-j})}, f\in \mathcal{F} \right\}, 
\end{align*}
are $P_0$-Glivenko-Cantelli with $j=1,\ldots,p$. 
Then, for any $\epsilon'>0$
\begin{align*}
\Pi \left\{ \max_{1\leq j \leq p}| \zeta_j(f,P_n) - \zeta_{0,j} | < \epsilon' \mid D_n \right\} \rightarrow 1, \ \ \text{{\it almost surely $P_0^{\infty}$}}.
%\label{eq:threshold}
\end{align*}
\end{theorem}

We illustrate a simple but non-trivial encompassing model which satisfies the sufficient conditions.  Let $y\in \Re$, $x\in \Re^p$ and $\phi_{\sigma}$ be the univariate normal density with mean 0 and standard deviation $\sigma$.  Then, we consider location mixtures of normals in which the kernel is  the product of a regression density for the response and independent normal densities for the predictors,
\begin{align}
f(y,x) &= \int \phi_{\sigma}(y - \tilde{x}^{\T}\beta) \prod_{j=1}^p \phi_{\tau_j}(x_j - \mu_j) Q(d\beta,d\mu), 
\label{eq:location2}
\end{align}
where $\tilde{x}=(1, x^{\T})^{\T}$, $\beta=(\beta_0,\ldots,\beta_p)^{\T}$, $\tau=(\tau_1,\ldots,\tau_p)^{\T}$ and $\mu=(\mu_1,\ldots,\mu_p)^{\T}$.  Dirichlet process mixture models of this type have been widely studied \citep{WestMullerEscobar94, EscobarWest95, MullerErkanliWest96, HannahBleiPowell11}.  We assume the mixing measure $Q$ can be expressed as
\begin{align}
Q = \sum_{h=1}^{\infty} \pi_h \delta_{(\beta_h,\mu_h)}, \ \ \pi_h \geq 0, \ \ \sum_{h=1}^{\infty} \pi_h = 1, \ \ (\beta_h,\mu_h) \sim G, 
\label{eq:Q2}
\end{align}
where $\beta_h=(\beta_{0,h},\ldots,\beta_{p,h})^{\T}$, $\mu_h=(\mu_{1,h},\ldots,\mu_{p,h})^{\T}$ and $G$ is a distribution on $\Re^{p+1}\times \Re^p$.  This class of functions (\ref{eq:location2}) and (\ref{eq:Q2}) includes Dirichlet process mixtures with $\pi_h=V_h\prod_{l<h}(1-V_l)$, $V_h\sim \text{Be}(1,\alpha_0)$ for $h=1,\ldots,\infty$ \citep{Sethuraman94}.  The prior distribution for the joint densities is induced through $\Pi=\Pi^{Q}\times \Pi^{(\sigma,\tau)}$ where $\Pi^Q$ and $\Pi^{(\sigma,\tau)}$ are the prior distributions for $Q$ and $(\sigma,\tau)$.  Under some conditions on $f_0$ and $\Pi$, the next lemma illustrates the encompassing model (\ref{eq:location2}) and (\ref{eq:Q2}) assures consistency.
\begin{lemma}
\label{lemma2}
Suppose the true density can be expressed in the form $f_0(y,x)=\int \phi_{\sigma_0}(y - \tilde{x}^{\T}\beta) \prod_{j=1}^p \phi_{\tau_{0,j}}(x_j - \mu_{j}) Q_0(d\beta,d\mu)$.  If $G$ has compact support, $\Pi^{(\sigma,\tau)}$ has compact support excluding zero, $Q_0$ belongs to the support of $\Pi^Q$ and ($\sigma_0$, $\tau_0$) are in the support of $\Pi^{(\sigma,\tau)}$, then $\Pi \left\{ \max_{1\leq j \leq p} | \zeta_j(f,P_n) - \zeta_{0,j} | < \epsilon' \mid D_n \right\} \rightarrow 1$ almost surely $P_0^{\infty}$.
\end{lemma}
The proof relies on Theorem 3 in \cite{GhosalGhoshRamamoorthi99} and is in the Supplementary
Material.  As Remark 1 in \cite{GhosalGhoshRamamoorthi99} mentions, the result can be extended to a wider class of location-scale mixture of normals.  The condition of compact support is sufficient but not necessary.

%%%%%%%%%%%%%%%%%%%%%%%%%%%%%%%%%%%%%%%%%%%%%%%%%%%%%%%%%%%%%%%%%%%%%%%%%%%%
\section{Simulation study} \label{simulation}
%%%%%%%%%%%%%%%%%%%%%%%%%%%%%%%%%%%%%%%%%%%%%%%%%%%%%%%%%%%%%%%%%%%%%%%%%%%%

In this section, we assess performance of the proposed method compared to frequentist nonparametric alternatives.  As competitors, we employ a method based on cumulative distribution functions with Cram\'{e}r-von-Mises type statistics from an unpublished 1996 technical report by O. Linton and P. Gozalo, the kernel measure method based on normalized cross-covariance operators on reproducing kernel Hilbert spaces \citep{FukumizuGrettonSun08} and the asymmetric quadratic measure \citep{SethPrincipe12}.  Matlab code for these methods is available at http://www.sohanseth.com/Home/codes and we use the default settings recommended in \cite{SethPrincipe12} with a Gaussian kernel for \cite{FukumizuGrettonSun08} and a Laplacian function for the asymmetric quadratic measure.  Also, for these methods, we reject the hypothesis $Y\perp X_j\mid X_{-j}$ if $ B^{-1}\sum_{b=1}^{B} 1(d^*_b>d)<0.1$ where $d$ and $d^*_b$ are the estimated conditional dependences using the observation and the $b$th randomly rearranged observation which mimics the case of conditional independence \citep{DiksDeGoede01} with $b=1,\ldots,B$ and $B=100$.  In addition, we apply the lasso function in Matlab using 5-fold cross validation for penalty coefficient selection and other default settings. We evaluate performance based on the following measures: type 1 error (false positive/(false positive+true negative)), type 2 error (false negative/(true positive+false negative)), positive predictive value (true positive/positive), negative predictive value (true negative/negative) and accuracy ((true positive+true negative)/(positive+negative)).  

As an encompassing model, we employ the following Dirichlet process location-scale mixture,
\begin{align}
f(y,x) &= \int \phi_{\sigma}(y - \tilde{x}^{\T}\beta ) \prod_{j=1}^p \phi_{\tau_j}(x_j-\mu_j) Q(d\beta, d\mu, d\sigma, d\tau ),  \label{eq:main-model} \\
&= \sum_{h=1}^{H} \pi_h \phi_{\sigma_{h}} (y - \tilde{x}^{\T}\beta_h ) \prod_{j=1}^p  \phi_{\tau_{j,h}}(x_j-\mu_{j,h}), 
\label{eq:main-model2}
\end{align}
where $\pi_{h}= V_h \prod_{l<h} (1 - V_l)$, $V_h \sim \text{Be}(1, \alpha_0)$ for $h=1,\ldots,H-1$ with $V_H = 1$, $\beta=(\beta_0,\ldots,\beta_p)^{\T}$, $\tilde{x}=(1, x^{\T})^{\T}$, $\mu=(\mu_1,\ldots,\mu_p)^{\T}$ and $\tau=(\tau_1,\ldots,\tau_p)^{\T}$.  As discussed in subsection \ref{variable selection}, if the base measure of the Dirichlet process has compact support, we obtain consistent estimators of the conditional mutual information for each predictor.  Compact support is a simplifying assumption for the theory, which can be relaxed, and we avoid this restriction in the computation letting $\sigma^2 \sim \text{Inverse-Gamma}(1.5, 0.5)$, $\mu_{j,h} \sim N(0,1)$, $\tau^2_{j,h} \sim \text{Inverse-Gamma}(1.5, 0.5)$ and $\alpha_0 \sim \text{Ga}(0.25,0.25)$. To allow a sparse regression structure, we use a point mass mixture prior: $\beta_j \sim p_0 \delta_0 + (1-p_0) N (0, \lambda^2_j)$, $\lambda^2_j \sim \text{Inverse-Gamma}(0.5, 0.5)$, $\lambda_j$ are mutually independent over $j$ and $p_0 \sim \text{Be}(4.75, 0.25)$.  By integrating out $\lambda^2_j$, this prior corresponds to a mixture of a degenerate distribution concentrated at zero and a Cauchy distribution.  The prior for exclusion probability $p_0$ assumes 5\% of regression coefficients out of $H(p+1)$ components are non-zero but allows substantial uncertainty since the prior sample size is set to be 4.75+0.25=5.  Also, we set $H=20$.  Before posterior computation, we normalize data to have mean zero and standard deviation one.  We draw 10,000 samples after the initial 5,000 samples are discarded as a burn-in period and every 10th sample is saved.  Rates of convergence and mixing were adequate.  Illustrative examples of sample paths and autocorrelations of $\zeta_j(f,P_n)$ are included in the Supplementary
Material.  We conclude there is substantial evidence of violations of $Y\perp X_j\mid X_{-j}$ if $\Pi\{ \zeta_j(f,P_n)>0\mid D_n\}>0.95$ with $j=1,\ldots,p$.

\begin{figure}
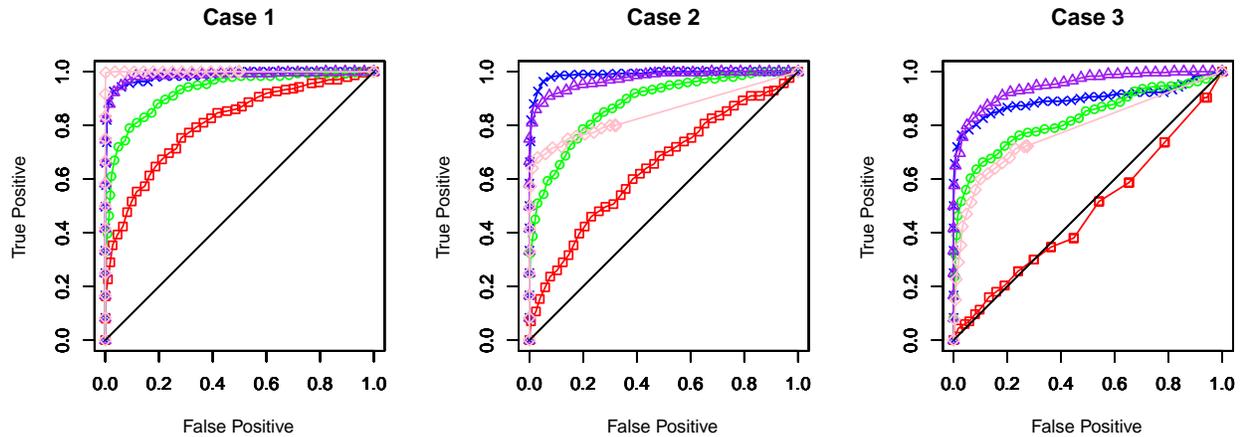

\vspace{-3cm}
\figurebox{35pc}{}{}[ROC-0303.eps]
\vspace{-3.5cm}
\caption{Receiver operating characteristic curves and area under the curve curves in Case 1 (left), Case 2 (middle) and Case 3 (right).  $y$ axis represents the true positive rate and $x$ axis the false positive rate.  Blue crosses, pink diamonds, red square, green circles and purple triangles indicate the averages of the true and false positive rates over 100 data sets for the proposed method, lasso, Cram\'{e}r-von-Mises type statistic, normalized cross-covariance operator and asymmetric quadratic measure.}
\label{fig:ROC1}
\end{figure}

We consider three different data-generating functions from which we simulate 100 data sets with $n=100$ and $p=10$.  First, we generate data from a linear regression model with strong dependence among predictors.
\begin{align*}
\text{Case 1}: \hspace{5mm}
y_i  &= - x_{i,1} + x_{i,4} - x_{i,7} + \varepsilon_i, \ \ \ \varepsilon_i \sim N(0, 1), \\
x_i &= (x_{i,1},\ldots,x_{i,10}) \sim N(0, \Sigma_x), \\
\Sigma_x &= \{\sigma_{j,j'}\}, \  
\sigma_{j,j'} = \text{cov}(x_{i,j},x_{i,j'})=0.7^{|j-j'|},
\end{align*}
where $\{y_i\}$ are independent over $i$. The left panel in Figure \ref{fig:ROC1} and last column in Table \ref{tb:sim} show the receiver operating characteristic curves and area under the curve averaged over 100 data sets in Case 1.  For the proposed method, we obtain the curve by shifting the threshold $a$ in $\Pi\{ \zeta_j(f,P_n)>a\mid D_n\}>0.95$.  For the lasso, we shift the threshold for absolute values of regression coefficients.  We set the thresholds as 2.5$k$\% quantile points of all estimated measures of conditional dependence over 100 data sets for each method with $k=0,\ldots,40$.  Although the area under the curve for the proposed method is slightly smaller than that for the lasso and the asymmetric quadratic measure, it is large and close to one.  The top of Table \ref{tb:sim} reports averaged measures of the test performance over 100 data sets in Case 1.  For the lasso, its high type 1 error and low positive predictive value indicate it incorrectly rejects many hypotheses.  Though the data are generated from the linear model, the strong dependence among predictors can cause poor performance.  On the other hand, high type 2 errors and low negative predictive values in the Cram\'{e}r-von-Mises type statistic and asymmetric quadratic measure imply that they often fail to detect dependent relations.  The normalized cross-covariance operator also faces the same problem of missing dependent predictors but the performance is much better.  The proposed method works quite well, reporting small type 1 and 2 errors and high positive  and negative predictive values.  Compared to the normalized cross-covariance operator, there is not a big difference in measures with false positives but the proposed method less often produces false negatives since the new approach shows a lower type 2 error and a higher negative predictive value.

Next, we generate data from a model in which the strong dependence among predictors remains but the relation between the response and predictors is non-linear.
\begin{align*}
\text{Case 2}: \hspace{5mm}
y_i  &= - x_{i,1} + \exp(x_{i,4}) - x_{i,7}^2 + \varepsilon_i, \ \ \ \varepsilon_i \sim N(0, 1), \\
x_i &= (x_{i,1},\ldots,x_{i,10}) \sim N(0, \Sigma_x), \\
\Sigma_x &= \{\sigma_{j,j'}\}, \  
\sigma_{j,j'} = \text{cov}(x_{i,j},x_{i,j'})=0.7^{|j-j'|}.
\end{align*}
The receiver operating characteristic curves and area under the curve in Case 2 are in the middle of Figure \ref{fig:ROC1} and Table \ref{tb:sim}.  Though the competitors' curves are away from the random guess line $y=x$, the proposed method shows largest area under the curve.  The middle of Table \ref{tb:sim} summarizes the test performance measures.  The proposed method reports small type 1 and 2 errors and high positive and negative predictive values and accuracy.  From the high type 1 error and small positive predictive value, the lasso tends to wrongly pick up conditionally independent predictors.  The high type 2 error and small negative predictive value indicate the Cram\'{e}r-von-Mises type statistic and asymmetric quadratic measure have difficulty in finding dependent structures.  The normalized cross-covariance operator performs better than the Cram\'{e}r-von-Mises type statistic and asymmetric quadratic measure but still reports a high type 2 error and a low negative predictive value compared to the proposed method.

We also simulate data from a different non-linear model where the dependence comes from division of the sample into subgroups and non-linear regressions.  
\begin{align*}
\text{Case 3}: \hspace{5mm}
& y_i  = 
\begin{cases}
             0.8 x^2_{i,1}   - x_{i,4} + \varepsilon_i, \ \ \ \varepsilon_i \sim N(0,0.7^2), & \text{if $s_i=0$}, \\
             - x_{i,1} + 1.2 \exp(x_{i,7}) + \varepsilon_i, \ \ \ \varepsilon_i \sim N(0,1), & \text{if $s_i=1$}.
\end{cases}    \\
& s_i \sim \text{Bernoulli}(0.5), \ \ 
 x_{i,j} \sim N(\mu_{j, s_i}, \sigma^2_{j, s_i}), \ j=1,\ldots,10, \\
& \mu_{j, s} \sim N(0, 1), \ \sigma^2_{j, s} \sim \text{Inverse-Gamma(2, 0.5)}, \ s\in\{0,1\}, \\
&\mu_{j, 0} = \mu_{j, 1}, \ \ \sigma^2_{j, 0} = \sigma^2_{j, 1}, \ \ j\notin \{1,4,7\}.
\end{align*}
The right plot in Figure \ref{fig:ROC1} and last column in Table \ref{tb:sim} correspond to the receiver operating characteristic curves and area under the curve in Case 3.  The Cram\'{e}r-von-Mises type statistic works poorly with the curve close to the random guess line.   The area under the curve by the proposed method is smaller than that for the asymmetric quadratic measure but the curve is still far away from the $y=x$ line.  The bottom in Table \ref{tb:sim} reports measures of the test performance.  The lasso is likely to reject correct hypotheses and the Cram\'{e}r-von-Mises type statistic produces the worst results in all measures except the type 1 error.  The proposed method, the normalized cross-covariance operator and asymmetric quadratic measure show small type 1 errors and high positive predictive values, indicating they less likely produce false positives.  As for the false negatives, the differences in the type 2 errors and negative predictive values between the proposed method and the normalized cross-covariance operator are small with the asymmetric quadratic measure slightly worse.  Also, the proposed method leads to the highest accuracy among them.  Overall these simulation results are promising that the proposed method has relatively good performance.

In addition, the proposed method can be applied for detecting marginal associations between two random variables by utilizing mutual information instead of conditional mutual information, that is, $\zeta(f,P_n)=\int f(y,x)/\{f(y)f(x)\} dP_n$.  We compared the proposed approach with \cite{Heller-et-al13} using the data $\{(y_i,x_i),i=1,\ldots,n\}$ from Case 1, 2 and 3 with an additional error, $y_i^*=y_i+\varepsilon_i^*$, $\varepsilon_i^*\sim N(0,\sigma^{*2})$.  For the competitor, we use R package HHG with default settings using 1,000 random permutations and 0.05 significance level. We observe the proposed method has better performance in detecting associations between $y^*_i$ and $x_i$ across $\sigma^*$ values.  We also find similar small type 1 error rates for the two methods in null settings.  The results are shown in the Supplementary Material.

\begin{table}[htp]
\tbl{Averages of type 1 and 2 errors, positive and negative predictive values, accuracy and area under the curve in Case 1 (top), Case 2 (middle) and Case 3 (bottom)}{%
\begin{tabular}{c|cccccccc}
%\\
\hline
Case 1	& Type 1	&	Type 2	& PPV & NPV  & ACC & AUC \\
\hline
Proposed 	&	2.2		&	12.6  & 95.5 & 95.6 & 94.6 & 98.4 \\
LASSO		&	49.7	&	0.0	 & 50.6 & 100.0 & 65.2 & 99.9 \\
CM			&	0.2		&	80.3 & 97.1 & 74.6 & 75.7 & 80.6 \\
NCCO		&	0.1		&	24.3 & 99.6 & 91.3 & 92.6 & 92.8 \\
AQM	 		&	0.0		&	67.6 & 100.0 & 77.9 & 79.7 & 98.6 \\
\hline
Case 2	& Type 1	&	Type 2	& PPV & NPV  & ACC & AUC \\
\hline
Proposed	&	4.0		&	12.0 & 92.8 & 95.5 & 93.6 & 98.9 \\
LASSO		&	32.0	&	20.0 & 58.5 & 89.1 & 71.6 & 84.8 \\
CM			&	1.7		&	90.6 & 71.9 & 71.7 & 71.6 & 64.3 \\
NCCO		&	0.2		&	37.0 & 99.4 & 87.4 & 88.7 & 87.8 \\
AQM		 	&	0.0		&	76.0 & 100.0 & 75.9 & 77.2 & 97.3 \\
\hline
Case 3	& Type 1	&	Type 2	& PPV & NPV  & ACC & AUC \\
\hline
Proposed	&	2.8		&	27.0 & 94.3 & 90.4 & 89.9 & 89.6 \\
LASSO		&	27.2	&	27.6 & 64.7 & 88.8 & 72.6 & 78.4 \\
CM			&	15.5	&	78.0 & 43.2 & 72.1 & 65.7 & 47.6 \\
NCCO		&	3.5		&	27.0 & 94.0 & 90.2 & 89.4 & 82.4 \\
AQM		 	&	0.2		&	41.3 & 99.4 & 85.5 & 87.4 & 94.7 \\
\hline
\end{tabular}}
\label{tb:sim}
\begin{tabnote}
Proposed, proposed method; CM, Cram\'{e}r-von-Mises type statistic; NCCO, normalized cross-covariance operator; AQM, asymmetric quadratic measure; PPV, positive predictive value; NPV, negative predictive value; ACC, accuracy; AUC, area under the curve.
\end{tabnote}
\end{table}

%%%%%%%%%%%%%%%%%%%%%%%%%%%%%%%%%%%%%%%%%%%%%%%%%%%%%%%%%%%%%%%%%%%%%%%%%%%%
\section{Application to criminology data} \label{application}
%%%%%%%%%%%%%%%%%%%%%%%%%%%%%%%%%%%%%%%%%%%%%%%%%%%%%%%%%%%%%%%%%%%%%%%%%%%%

In this section, we apply the proposed method to communities and crime data from the University of California Irvine machine learning repository.  Details of the data are in the Supplementary Material.  The data set is culled from 1990 United States census, 1995 United States Federal Bureau of Investigation uniform crime report and 1990 United States law enforcement management and administrative statistics survey.  Data include various types of crime and demographic information for $n=2,215$ communities in the United States.  We use 10 count variables as responses: numbers of murders, rapes, robberies, assaults, burglaries, larcenies, auto thefts, arsons, violent crimes (sum of murders, rapes, robberies and assaults) and non-violent crimes (sum of burglaries, larcenies, auto thefts and arsons).  As predictors, we select $p=68$ variables, such as per capita income and population density, which indicate demographic characteristics of the communities.  The list is in the Supplementary Material.  The data set consists of count, percentage and positive continuous variables.  We observe the count variables have right-skewed distributions and the percentage variables can inflate at 0\% and 100\%.  Also, the data set includes missing values in the response.

To incorporate mixed-scale measurements, we develop a joint model which relies on the rounded kernel method of \cite{CanaleDunson11}.  Let $y^*\in \Re$ and $x^*=(x_1^*,\ldots,x_p^*)^{\T}\in \Re^{p}$ be latent continuous variables for the response $y$ and predictors $x=(x_1,\ldots,x_p)^{\T}$.  We induce a flexible nonparametric model on $y$ and $x$ through a Dirichlet process mixture of normals for the latent variables.  If $x_j$ is a count variable, it can be expressed as $x_j=l$ if $a_l<x^*_j\leq a_{l+1}$ with $l=0,1,2,\ldots$ where $-\infty=a_0<a_1<a_2<\cdots$ with $a_l=\log(l)$ for $l \geq 1$.  This expression corresponds to $x_j = [\exp(x_j^*)]$ where $[x]$ denotes the maximum integer smaller than $x$.  Since the log function shrinks large values, a distribution with positive skewness can be efficiently approximated by mixtures of normals with the log cut-points.  Percentage variables with inflation at 0 and 100\% can be induced by 
\begin{align*}
x_j = \begin{cases}
    0 & \text{if $x_j^* \leq 0$}, \\
    x_j^*  & \text{if $0 < x_j^* < 100$}, \\
    100 & \text{if $100 \leq x_j^*.$}
  \end{cases}
\end{align*}
As for a positive continuous variable, we apply the log transformation to the original data and treat it as a continuous variable with $x_j=x_j^*$.  For the latent variables, we utilize the Dirichlet process mixture of normals (\ref{eq:main-model}) and (\ref{eq:main-model2}) except we use the observed predictors for the regression on $y^*$.  Then, we obtain the following joint model of $y$ and $x$ by integrating out the latent variables.
\begin{align}
f(y,x) = \sum_{h=1}^{H} \pi_h f(y \mid x, \theta_h) \prod_{j=1}^p f(x_{j} \mid \theta_h), \label{eq:joint-density}
\end{align}
where $\pi_{h}= V_h \prod_{l<h} (1 - V_l)$, $V_h \sim \text{Be}(1, \alpha_0)$ for $h=1,\ldots,H-1$ with $V_H = 1$, $\theta$ is a parameter set in the model and   
\begin{align}
f(y \mid x, \theta) &= \int_{a_{y}}^{a_{y+1}} \phi_{\sigma}(y^* - \tilde{x}^{\T} \beta) d y^* = \Phi(a_{y+1} \mid \tilde{x}^{\T} \beta, \sigma) - \Phi(a_{y} \mid \tilde{x}^{\T} \beta, \sigma), \label{eq:standardize}
\end{align}
and
\begin{align*}
f(x_{j} \mid \theta) = \begin{cases}
\text{count:} \ \ \Phi(a_{x_{j}+1} \mid \mu_{j}, \tau_{j}) - \Phi(a_{x_{j}} \mid \mu_{j}, \tau_{j}), \\
\text{percentage:} \ \ 1(x_{j}=0)\Phi(0 \mid \mu_{j}, \tau_{j}) + 1(x_{j}=100) \{ 1-\Phi(100 \mid \mu_{j}, \tau_{j}) \}  \\
\ \ + 1(0<x_{j}<100) \phi_{\tau_j}(x_{j} - \mu_{j}), \\
\text{continuous:} \ \ \phi_{\tau_j}(x_{j} - \mu_{j}),
\end{cases}
\end{align*}
where $1(\cdot)$ is an indicator function and $\Phi(\cdot \,|\, a, b)$ is the cumulative density function of normal with mean $a$ and standard deviation $b$.  We constructed priors relying on empirical information, $\sigma^2 \sim \text{Inverse-Gamma}(1.5, s^2_y/2)$ where $s^2_y$ is the sample variance of $\log(y_i+0.5)$ since $y_i=0$  for certain subjects.  Also, we use $\mu_{j}\sim N(\bar{\mu}_j,s^2_j)$ and $\tau_{j}^2 \sim \text{Inverse-Gamma}(1.5, s^2_j/2)$ where $\bar{\mu}_j$ and $s^2_j$ are the sample mean and variance of $\log(x_{i,j}+0.5)$ for a count and of $x_{i,j}$ for a percentage and a continuous variable.  The priors for $\alpha_0$ and $\beta$ are the same as in Section \ref{simulation}.  We standardize the predictors in (\ref{eq:standardize}) so that each variable has mean zero and standard deviation one.  Assuming missing at random, we impute missing values at each Markov chain Monte Carlo iteration from the conditional distributions given observed data.  The details of the Markov chain Monte Carlo algorithm are in the Supplementary Material.  We apply the proposed method with $H=20$ separately to each response.  We draw 80,000 samples from the posterior after the initial 5,000 samples are discarded as a burn-in period and every 20th sample is saved.  We observe that the sample paths were stable and the sample autocorrelations dropped smoothly; hence we concluded the chains converged.  The sample paths and autocorrelations of $\zeta_j(f,P_n)$ with several $j$ for each response are in the Supplementary Material.  In the computation of $\zeta_j(f,P_n)$, we need to evaluate $f(y_i,x_{i,-j})$ but it is not straightforward to integrate $x_j$ out from the joint density (\ref{eq:joint-density}).  Hence, we apply a Monte Carlo approximation based on 500 random samples from $f(x_{i,j}\mid \theta_h)$ for each $h$.

\begin{figure}[htp]
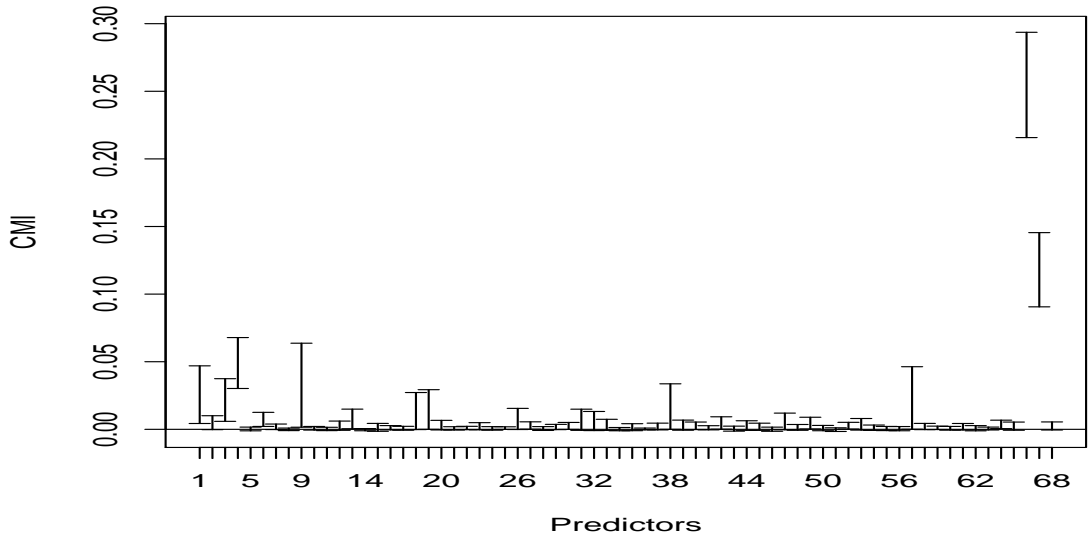

\figurebox{20pc}{37pc}{}[Plot-murder.eps]
\caption{90\% credible intervals of the estimated conditional mutual information with murder as the response for each of the 68 demographic predictors adjusting for the other predictors.}
\label{fig:select1}
\end{figure}

Figure \ref{fig:select1} shows 90\% credible intervals of $\zeta_j(f,P_n)$ for all $j$ and Table \ref{tb:list1} reports the top 10 selected predictors in descending order of the posterior mean of conditional mutual information for murders.  Full lists of the selected predictors for all responses are in the Supplementary Material.  Certain predictors are selected for many different crime-related response variables.  For all crimes, land area and population density show the first and second largest conditional dependence adjusting for other factors.  Also, their posterior means of the conditional mutual information are much larger than those of other predictors especially in burglaries, larcenies, auto thefts and non-violent crimes.  In addition, population in urban areas is selected 8 times, population, the percentage of kids with two parents and the percentage of persons in dense housing are picked up 7 times, and the percentage of Caucasian, the percentage of households with investment and rent income, the percentage of housing occupied and the percentage of families with two parents are conditionally dependent with 6 types of crimes.  On the other hand, 12 predictors such as the percentage of housing units with less than 3 bedrooms and the percentage of moms of kids under 18 in labor force are not selected for any crimes.

Also, we can find similarities in the top 10 selected predictors among all crimes.  We observe that certain types of variables obtain high ranks for many responses.  For example, all crimes except larcenies and auto thefts share at least one of population in the community and population in urban areas in their lists.  In addition, the percentage of families with parents and the percentage of kids with parents show relatively strong conditional dependence with all crimes other than murders, auto thefts and arsons.  The posterior means of conditional mutual information of race variables are large for murders, robberies, assaults and violent crimes.  Also, the top 10 lists of rapes, burglaries, arsons and non-violent crimes include more than one predictor related to divorce.

We also apply the competitors discussed in Section \ref{simulation} to the crime data using the same default settings.  For the missing values, we impute them by the mean of observed values.  The lists of the selected predictors are in the Supplementary Material.  The Cram\'{e}r-von-Mises type statistic seems to work poorly in that it selects all predictors for all crimes.  The predictors selected by the lasso are overlapping with those by the proposed method, such as population and the percentage of housing occupied, but the land areas and population density are often missed.  The normalized cross-covariance operator shows little difference over crimes.  It basically selects the same sets of predictors for all crimes but the land area and population density are not included.  The asymmetric quadratic measure shares some predictors such as race with the proposed method but fails to pick up the top 2 variables as well.  The inability of the other methods to detect these important predictors is likely due to their non-linear and non-monotonic relationship with the crime responses.

\begin{table}[htp]
\tbl{Top 10 selected predictors in descending order of the posterior means of conditional mutual information with murders as the response.}{%
\begin{tabular}{cccl}
\hline
\multicolumn{1}{c}{$j$} & \multicolumn{1}{c}{Mean}	& \multicolumn{1}{c}{90\%CI}	& \multicolumn{1}{l}{Predictor}	\\
\hline 
66 & 0.2587 & [0.2157, 0.2936] & land area in square miles \\
67 & 0.1188 & [0.0905, 0.1454] & population density in persons per square mile \\ 
4  & 0.0507 & [0.0302, 0.0678] & \% of population that is caucasian  \\ 
9  & 0.0250 & [0.0043, 0.0636] & \# of people living in areas classified as urban \\ 
1  & 0.0250 & [0.0015, 0.0469] & population for community  \\ 
3  & 0.0192 & [0.0058, 0.0374] & \% of population that is african american \\ 
57 & 0.0177 & [0.00007, 0.0463] & rental housing: lower quartile rent \\ 
13 & 0.0075 & [0.0004, 0.0149] & \% of households with investment / rent income in 1989 \\ 
6  & 0.0067 & [0.0021, 0.0125] & \% of population that is of hispanic heritage  \\ 
64 & 0.0039 & [0.0005, 0.0067] & \% of people born in the same state as currently living  \\ 
\hline
\end{tabular}}
\label{tb:list1}
\begin{tabnote}
$j$, $j$-th predictor; Mean, posterior mean; 90\% CI corresponds to a 90\% credible interval.
\end{tabnote}
\end{table}

\section*{Acknowledgement}
This work was supported by Nakajima Foundation and grants  from the National Institute of Environmental Health Sciences of the United States National Institutes of Health.  The computational results are mainly generated using Ox \citep{Doornik06} and Matlab.

\section*{Supplementary material}

Supplementary material available at Biometrika online includes proofs of Theorem 2 and Lemma 1, details of the data set, the Markov chain Monte Carlo algorithm and additional estimation results in Section \ref{simulation} and \ref{application}.

\appendix

\appendixone
\section*{Appendix 1} \label{ap:theorem}
\subsection*{Proof of Theorem 1} 

For $\epsilon>0$, we define $E=[ f: KL\{f_0(y,x,z),f(y,x,z)\}<\epsilon]$.  Then, there exists $N$ such that for $n>N$ and $f\in E$,
\begin{align}
& |\zeta(f,P_n) - \zeta_0| = \left| \int \log \frac{f(y,x,z) f(z)}{f(y,z)f(x,z)} dP_n - \int \log \frac{f_0(y,x,z) f_0(z)}{f_0(y,z)f_0(x,z)} dP_0  \right|, \nonumber \\
& \leq \sup_{f\in\mathcal{F}} \bigg|  \int \log \frac{f_0(y,x,z)}{f(y,x,z)} dP_n - \int \log \frac{f_0(y,x,z)}{f(y,x,z)} dP_0 \bigg| 
+ \sup_{f\in\mathcal{F}} \bigg|  \int \log \frac{f_0(y,z)}{f(y,z)} dP_n - \int \log \frac{f_0(y,z)}{f(y,z)} dP_0 \bigg| 
\label{eq:gc1}  \\
& \ + \sup_{f\in\mathcal{F}} \bigg|  \int \log \frac{f_0(x,z)}{f(x,z)} dP_n - \int \log \frac{f_0(x,z)}{f(x,z)} dP_0 \bigg| 
+ \sup_{f\in\mathcal{F}} \bigg|  \int \log \frac{f_0(z)}{f(z)} dP_n - \int \log \frac{f_0(z)}{f(z)} dP_0 \bigg|  
\label{eq:gc4} \\
& \ + \left| \int \log \frac{f_0(y,x,z) f_0(z)}{f_0(y,z)f_0(x,z)} dP_n - \int \log \frac{f_0(y,x,z) f_0(z)}{f_0(y,z)f_0(x,z)} dP_0  \right|  
+ \int \log \frac{f_0(y,x,z)}{f(y,x,z)} dP_0 + \int \log \frac{f_0(y,z)}{f(y,z)} dP_0 
\label{eq:kls} \\
& \ + \int \log \frac{f_0(x,z)}{f(x,z)} dP_0
+ \int \log \frac{f_0(z)}{f(z)} dP_0 
\leq 9 \epsilon, \ \  \text{almost surely.} \label{eq:kls2}
\end{align}
Each term in (\ref{eq:gc1})-(\ref{eq:gc4}) can be bounded by $\epsilon$ almost surely from the definition of $P_0$-Glivenko-Cantelli classes.  The first term in (\ref{eq:kls}) goes to zero by the strong law of large numbers. The other terms in (\ref{eq:kls}) and the terms in (\ref{eq:kls2}) are bounded by $2\epsilon$ almost surely respectively.  This comes from the non-negativity of the Kullback-Leibler divergence, for example,
\begin{align*}
\int \log \frac{f_0(y,z)}{f(y,z)} dP_0 &\leq \int \log \frac{f_0(y,z)}{f(y,z)} dP_0 + \int \log \frac{f_0(x\mid y,z)}{f(x\mid y,z)} dP_0  = \int \log \frac{f_0(y,x,z)}{f(y,x,z)} dP_0 < \epsilon.
\end{align*}
Hence, by setting $\epsilon'=9\epsilon$, $E\subset \{f: |\zeta(f,P_n) - \zeta_0| < \epsilon' \}$.  The argument by A. Norets in the Supplementary Material shows if $[\log\{f_0(y,x,z)/f(y,x,z)\}, f\in \mathcal{F}]$ is $P_0$-Glivenko-Cantelli and the Kullback-Leibler support condition (\ref{eq:kl-support}) is satisfied, then the posterior converges to the true data-generating function in the Kullback-Leibler distance.  Therefore, $\Pi\{ |\zeta(f,P_n) - \zeta_0| < \epsilon' \mid D_n \} \geq \Pi(E\mid D_n)\rightarrow 1$ almost surely $P_0^{\infty}$.

\bibliographystyle{biometrika}
\bibliography{cmi2}

\end{document}

\jname{Biometrika}
%% The year, volume, and number are determined on publication
%\jyear{}
%\jvol{}
%\jnum{}
%% The \doi{...} and \accessdate commands are used by the production team
%\doi{10.1093/biomet/asm023}
\accessdate{}
\copyrightinfo{\Copyright\ 2014 Biometrika Trust\goodbreak {\em Printed in Great Britain}}

%% These dates are usually set by the production team
\received{August 2014}
%\revised{September 2012}

%% The left and right page headers are defined here:
\markboth{T. Kunihama \and D. B. Dunson}{Biometrika style}

%% Here are the title, author names and addresses
\title{Supplementary material for Nonparametric Bayes inference on conditional independence}

\author{T. Kunihama \and D. B. Dunson}
\affil{Department of Statistical Science, Duke University, Durham, NC 27708-0251, USA \email{tsuyoshi.kunihama@duke.edu} \email{dunson@duke.edu}}

\maketitle

\section{Posterior consistency for $P_0$-Glivenko-Cantelli class}

{\bf Argument} (A. Norets)
{\it
Suppose $\{\log(f_0/f),f\in\mathcal{F}\}$ is a $P_0$-Glivenko-Cantelli class of functions and for any $\epsilon>0$,
\begin{align}
\Pi \left\{ KL(f_0,f) < \epsilon  \right\} > 0.
\label{eq:kl-support2}
\end{align}
Then, for any $\epsilon'>0$ and $E=\{f: KL(f_0,f) < \epsilon' \}$,
\begin{align*}
\Pi ( E^c \mid D_n) \rightarrow 0, \ \ \ \ \text{almost surely $P_0^{\infty}$}.
\end{align*}
}

\begin{proof} This proof is from a 2012 unpublished technical paper of A. Norets. The posterior can be expressed as
\begin{align*}
\Pi(E^c\mid D_n) &= \frac{\int_{E^c} \prod_{i=1}^n f(x_i)/f_0(x_i) d\Pi(f) }{ \int_{\mathcal{F}} \prod_{i=1}^n f(x_i)/f_0(x_i) d\Pi(f)}, \\
&= \frac{\exp(n\epsilon/2) \int_{E^c} \exp[ \sum_{i=1}^n \log\{ f(x_i) / f_0(x_i) \} ] d\Pi(f)}{ \exp(n\epsilon/2) \int_{\mathcal{F}} \exp[ \sum_{i=1}^n \log\{ f(x_i) / f_0(x_i) \}   d\Pi(f)}.
\end{align*}
The numerator can be expressed as
\begin{align*}
& \int_{KL(f_0,f)\geq \epsilon} \exp \left[ n\left\{ \frac{\epsilon}{2} - KL(f_0,f) + KL(f_0,f) - \frac{1}{n}\sum_{i=1}^n \log \frac{f_0(x_i)}{f(x_i)} \right\} \right] d\Pi(f), \\ 
& \leq \exp \left[ -n \left\{ \frac{\epsilon}{2} - \sup_{f\in\mathcal{F}} \left| \frac{1}{n}\sum_{i=1}^n \log \frac{f_0(x_i)}{f(x_i)} - \int \log \frac{f_0(x)}{f(x)} dP_0 \right| \right\} \right] \rightarrow 0
\end{align*}
almost surely $P^{\infty}_0$ because $\{\log(f_0/f),f\in\mathcal{F}\}$ is a $P_0$-Glivenko-Cantelli class.  Also, the denominator can be bounded below by
\begin{align*}
& \int_{KL(f_0,f)< \epsilon/4} \exp \left[ n\left\{ \frac{\epsilon}{2} - KL(f_0,f) + KL(f_0,f) - \frac{1}{n}\sum_{i=1}^n \log \frac{f_0(x_i)}{f(x_i)} \right\} \right] d\Pi(f), \\ 
& \geq \Pi\{KL(f_0,f)< \epsilon/4 \} \exp \left[ n \left\{ \frac{\epsilon}{4} - \sup_{f\in\mathcal{F}} \left| \frac{1}{n}\sum_{i=1}^n \log \frac{f_0(x_i)}{f(x_i)} - \int \log \frac{f_0(x)}{f(x)} dP_0 \right| \right\} \right] \rightarrow \infty
\end{align*}
from the assumption that $\Pi$ satisfies the $KL$ support condition and $\{\log(f_0/f),f\in\mathcal{F}\}$ is a $P_0$-Glivenko-Cantelli class.  Therefore, $\Pi ( E^c \mid D_n) \rightarrow 0$ almost surely $P_0^{\infty}$.
\end{proof}

\section{Proof of Theorem 2} \label{ap:theorem2}

For $\epsilon>0$, we define $E=[f: KL\{f_0(y,x),f(y,x)\}<\epsilon]$.  Then, there exists $N$ such that for $n>N$ and $f\in E$,
\begin{align}
\max_{1\leq j \leq p} |\zeta_j(f,P_n) - \zeta_0| &= \max_{1\leq j \leq p} \left| \int \log \frac{f(y,x) f(x_{-j})}{f(y,x_{-j})f(x)} dP_n - \int \log \frac{f_0(y,x) f_0(x_{-j})}{f_0(y,x_{-j})f_0(x)} dP_0  \right|, \nonumber \\
& \leq \sup_{f\in\mathcal{F}} \bigg|  \int \log \frac{f_0(y,x)}{f(y,x)} dP_n - \int \log \frac{f_0(y,x)}{f(y,x)} dP_0 \bigg|  \label{eq:gc1-2}  \\
&+ \max_{1\leq j \leq p} \sup_{f\in\mathcal{F}} \bigg|  \int \log \frac{f_0(y,x_{-j})}{f(y,x_{-j})} dP_n - \int \log \frac{f_0(y,x_{-j})}{f(y,x_{-j})} dP_0 \bigg| \\
&+ \sup_{f\in\mathcal{F}} \bigg|  \int \log \frac{f_0(x)}{f(x)} dP_n - \int \log \frac{f_0(x)}{f(x)} dP_0 \bigg| \\
&+ \max_{1\leq j \leq p} \sup_{f\in\mathcal{F}} \bigg|  \int \log \frac{f_0(x_{-j})}{f(x_{-j})} dP_n - \int \log \frac{f_0(x_{-j})}{f(x_{-j})} dP_0 \bigg|  
\label{eq:gc4-2} \\
&+ \max_{1\leq j \leq p} \left| \int \log \frac{f_0(y,x) f_0(x_{-j})}{f_0(y,x_{-j})f_0(x)} dP_n - \int \log \frac{f_0(y,x) f_0(x_{-j})}{f_0(y,x_{-j})f_0(x)} dP_0  \right|  \label{eq:lln-2} \\
&+ \int \log \frac{f_0(y,x)}{f(y,x)} dP_0 + \max_{1\leq j \leq p} \int \log \frac{f_0(y,x_{-j})}{f(y,x_{-j})} dP_0 \label{eq:kls-2} \\
&+ \int \log \frac{f_0(x)}{f(x)} dP_0
+ \max_{1\leq j \leq p} \int \log \frac{f_0(x_{-j})}{f(x_{-j})} dP_0, \label{eq:kls2-2} \\ 
&\leq 9 \epsilon, \ \ \text{almost surely}. \nonumber
\end{align}
(\ref{eq:gc1-2})-(\ref{eq:gc4-2}) are less than $\epsilon$ almost surely from the definition of $P_0$-Glivenko-Cantelli classes.  (\ref{eq:lln-2}) converges to zero by the strong law of large numbers.  Each term in (\ref{eq:kls-2}) and (\ref{eq:kls2-2}) are bounded by $KL\{f_0(y,x),f(y,x)\}$, which is less than $\epsilon$ almost surely.  Therefore, $E\subset \{f: \max_{1 \leq j \leq p} |\zeta_j(f,P_n) - \zeta_0| < \epsilon' \}$ where $\epsilon'=9\epsilon$ and $\Pi\{ \max_{1 \leq j \leq p} |\zeta_j(f,P_n) - \zeta_0| < \epsilon' \mid D_n\} \geq \Pi(E\mid D_n)\rightarrow 1$ almost surely $P_0^{\infty}$ from the posterior consistency of the joint densities in Kullback-Leibler divergence from the argument by A. Norets.

\section{Proof of Lemma 1} \label{ap:proposition2}

Without loss of generality, we assume $p=2$ and $\beta_0=0$.  
We first show that the Kullback-Leibler support condition holds for the encompassing model.  Since $Q_0$ and $G$ have compact support, we suppose $Q_0(A)=1$ and $Q(B)=1$ for $Q$ in the support of $\Pi^{Q}$ where  $A=\{(\beta,\mu): -k \leq \beta_1,\beta_2, \mu_1, \mu_2 \leq k\}$ and $B=\{(\beta,\mu): -k' \leq \beta_1,\beta_2, \mu_1, \mu_2 \leq k'\}$.  We can check $f_0$ has moments of all orders.  Hence, for any $\eta > 0$, there exists $a$ such that $\int_{|y|>a}  g(y,x) f_0(y,x)dydx  < \eta$, $\int_{|x_1|>a}  g(y,x) f_0(y,x)dydx < \eta$ and $\int_{|x_2|>a}  g(y,x) f_0(y,x)dydx < \eta$ where $g(y,x)=1+|x_1|+|x_2|+x_1^2+x_2^2+|y||x_1|+|y||x_2|+|x_1||x_2|$.  The Kullback-Leibler divergence between $f_0$ and $f$ can be expressed as
\begin{align}
\int f_0 \log \frac{f_0}{f} &= \int f_0(y,x) \log \frac{\int \phi_{\sigma_{0}}(y-x^{\T}\beta)\phi_{\tau_{0,1}}(x_1-\mu_1)\phi_{\tau_{0,2}}(x_2-\mu_2) dQ_0(\beta,\mu)}{\int \phi_{\sigma}(y-x^{\T}\beta)\phi_{\tau_{1}}(x_1-\mu_1)\phi_{\tau_{2}}(x_2-\mu_2) dQ_0(\beta,\mu)} dy dx \label{eq:right2-1} \\
&+ \int f_0(y,x) \log \frac{\int \phi_{\sigma}(y-x^{\T}\beta)\phi_{\tau_{1}}(x_1-\mu_1)\phi_{\tau_{2}}(x_2-\mu_2) dQ_0(\beta,\mu)}{\int \phi_{\sigma}(y-x^{\T}\beta)\phi_{\tau_{1}}(x_1-\mu_1)\phi_{\tau_{2}}(x_2-\mu_2) dQ(\beta,\mu)} dy dx. \label{eq:right2-2}
\end{align}
With respect to the integral (\ref{eq:right2-2}), we divide the support $\mathcal{R}^3$ into $C=\{(y,x)\in \mathcal{R}^3:-a\leq y,x_1,x_2 \leq a\}$ and its complement $C^C$.  For the complement, we consider the subspace $\{(y,x)\in \mathcal{R}^3: y< -a,-a\leq x_1,x_2\leq a\}$ for example.
\begin{align}
&\int_{-\infty}^{-a}  \int_{-a}^a \int_{-a}^a f_0(y,x) \log \frac{\int \phi_{\sigma}(y-x^{\T}\beta) \phi_{\tau_{1}}(x_1-\mu_1)\phi_{\tau_{2}}(x_2-\mu_2) dQ_0(\beta,\mu)}{\int \phi_{\sigma}(y-x^{\T}\beta) \phi_{\tau_{1}}(x_1-\mu_1)\phi_{\tau_{2}}(x_2-\mu_2) dQ(\beta,\mu)} dy dx, \nonumber \\
& \int_{-\infty}^{-a}  \int_{-a}^a \int_{-a}^a f_0(y,x) \log \frac{ \sup_{(\beta,\mu)\in A} \phi_{\sigma}(y-x^{\T}\beta) \phi_{\tau_{1}}(x_1-\mu_1)\phi_{\tau_{2}}(x_2-\mu_2) }{\inf_{(\beta,\mu)\in B} \phi_{\sigma}(y-x^{\T}\beta) \phi_{\tau_{1}}(x_1-\mu_1)\phi_{\tau_{2}}(x_2-\mu_2) } dy dx, \nonumber \\
& \leq \int_{-\infty}^{-a} \int \int \frac{1}{2\sigma^2} \{ (k^2+k'^2)(x_1^2+x_2^2) + 2(k+k')(|x_1|+|x_2|)|y| + 2(k^2+k'^2)|x_1||x_2| \} \nonumber \\
&\times f_0(y,x) dydx \nonumber \\
&+ \int_{-\infty}^{-a} \int \int  \left( \frac{k+k'}{\tau_1^2} |x_1| + \frac{k^2+k'^2}{2\tau_1^2} + \frac{k+k'}{\tau_2^2} |x_2| + \frac{k^2+k'^2}{2\tau_2^2} \right) f_0(y,x) dy dx \nonumber \\
&< \left( \frac{k+k'}{\sigma^2} + \frac{3(k^2+k'^2)}{2\sigma^2} + \frac{k+k'}{\tau_1^2} + \frac{k^2+k'^2}{2\tau_1^2} + \frac{k+k'}{\tau_2^2} + \frac{k^2+k'^2}{2\tau_2^2} \right) \eta. 
\label{eq:upperbound2-1}
\end{align}
For other regions in $C^C$ where one of $y$, $x_1$ and $x_2$ is larger than $a$ or smaller than $-a$, the corresponding integral can be bounded by (\ref{eq:upperbound2-1}).  Following the proof of Theorem 3 in \cite{GhosalGhoshRamamoorthi99}, there exists a set $E$ with $\Pi^{Q}(E)>0$ and for $Q\in E$, the integral over $C$ is less than $3\tilde{\eta}/(1-3\tilde{\eta})$ where $0<\tilde{\eta}<1/3$.  Therefore, for $Q\in E$, the integral (\ref{eq:right2-2}) is less than
\begin{align*}
6 \left( \frac{k+k'}{\sigma^2} + \frac{3(k^2+k'^2)}{2\sigma^2} + \frac{k+k'}{\tau_1^2} + \frac{k^2+k'^2}{2\tau_1^2} + \frac{k+k'}{\tau_2^2} + \frac{k^2+k'^2}{2\tau_2^2} \right) \eta + \frac{3\tilde{\eta}}{1-3 \tilde{\eta}}. 
%\label{eq:upperbound2-2}
\end{align*}

Also, we can show the right term in (\ref{eq:right2-1}) converges to 0 as $\sigma \rightarrow \sigma_{0}$, $\tau_j \rightarrow \tau_{0,j}$ with $j=1,2$ by the dominated convergence theorem with the inequality
\begin{align*}
\frac{\int \phi_{\sigma_{0}}(y-x^{\T}\beta)\phi_{\tau_{0,1}}(x_1-\mu_1)\phi_{\tau_{0,2}}(x_2-\mu_2) dQ_0(\beta,\mu)}{\int \phi_{\sigma}(y-x^{\T}\beta)\phi_{\tau_{1}}(x_1-\mu_1)\phi_{\tau_{2}}(x_2-\mu_2) dQ_0(\beta,\mu)}, \\ 
\leq \sup_{(\beta,\mu) \in A} \frac{\phi_{\sigma_{0}}(y-x^{\T}\beta)\phi_{\tau_{0,1}}(x_1-\mu_1)\phi_{\tau_{0,2}}(x_2-\mu_2)}{ \phi_{\sigma}(y-x^{\T}\beta)\phi_{\tau_{1}}(x_1-\mu_1)\phi_{\tau_{2}}(x_2-\mu_2)}.
\end{align*}
For any $\epsilon>0$, we can choose $\eta$, $\tilde{\eta}$ and a small neighborhood of $\sigma_0$ and $\tau_0$ such that both the integrals in (\ref{eq:right2-1}) and (\ref{eq:right2-2}) are less than $\epsilon/2$ respectively.  Then, the Kullback-Leibler support condition is satisfied.

Next, we check the Glivenko-Cantelli conditions.  For simplicity, we show only $[\log\{f_0(x_1)/f(x_1)\},f \in \mathcal{F}]$ is $P_0$-Glivenko-Cantelli but we can similarly prove that other classes of functions also satisfy the condition.  According to Theorem 3 in \cite{vanderVaartWellner00}, if two classes of functions $\mathcal{F}_0$ and $\mathcal{F}_1$ are $P_0$-Glivenko-Cantelli, then $g(\mathcal{F}_0,\mathcal{F}_1)$ is also $P_0$-Glivenko-Cantelli with $g$ a continuous function provided that it has an integrable envelope function.  We set $\mathcal{F}_0=\{f_0(x_1)\}$, $\mathcal{F}_1=\{f(x_1),f\in\mathcal{F}\}$ and $g$ is a log ratio function.  It is clear $\mathcal{F}_0$ is $P_0$-Glivenko-Cantelli.  Then, we show $\mathcal{F}_1$ is $P_0$-Glivenko-Cantelli by proving $\mathcal{F}_1$ satisfies the sufficient condition, $N_{[]}\{\epsilon,\mathcal{F}_1, L_1(P_0)\}<\infty$ for any $\epsilon>0$ where $N_{[]}\{\epsilon,\mathcal{F}_1, L_1(P_0)\}$ is the minimum number of $\epsilon$-brackets with which $\mathcal{F}_1$ can be covered in $L_1(P_0)$ distance.

We first construct bracket functions.  Let $[\underline{\tau}, \overline{\tau}]$ be the support of $\tau_1$.  Because the support of $(\mu_1,\tau_1)$ is compact, for any $\epsilon>0$ we can take $h>0$ such that $f(x_1)=\int\phi_{\tau_1}(x_1-\mu_1)dQ(\mu_1)<\epsilon$ for $|x_1|>h$ and any $\tau_1\in[\underline{\tau}, \overline{\tau}]$.  Also, we can show that $|f'(x_1)| < K$ for $x_1\in[-h,h]$ with some constant $K$.  Then, we take $0<\epsilon'<\epsilon/(K+1)$ and divide the interval $[-h,h]$ into sub-intervals $\{I_i,i=1,\ldots,G\}$ of equal length less than $\epsilon'$ with $[-h,h]=\cup_{i}I_i$ and $I_i\cap I_j=\emptyset$ for $i\neq j$.  On each interval $I_i$, we define $u_{ij}=(j\epsilon'+\epsilon)1_{I_i}$ and $l_{ij}=(j\epsilon')1_{I_i}$ for $j=0,\ldots,J$ such that $J\epsilon'>\max_{x_1\in[-h,h]} \max_{\tau_1\in[\underline{\tau},\overline{\tau}]} f(x_1)$ where $1_{I}$ is an indicator function on the interval $I$.  Letting $m_i\in \{1,\ldots,J\}$ and $m=(m_1,\ldots,m_G)$, we define $u_{m}=\sum_{i=1}^{G}u_{im_i}+\epsilon 1_{[-h,h]^C}$ and $l_m=\sum_{i=1}^G l_{im_i}$.  Then, it is straightforward to check $l_m<u_m$ and $||u_m-l_m||_{L_1(P_0)}\leq ||u_m-l_m||_{\infty}<\epsilon$.  Because $|f'(x_1)|<K$ and $\epsilon/\epsilon'>K+1$, for any $f\in\mathcal{F}_1$ there exists $m_i$ such that $l_{im_i}\leq f \leq u_{im_i}$ on the interval $I_i$ and further we can find some $m$ such that $l_m \leq f \leq u_m$ on $\mathcal{R}$.  Since $m\in\{1,\ldots,J\}^G$, the set $\{(l_m,u_m)\}$ consists of a finite number of functions.  Therefore, $N_{[]}\{\epsilon,\mathcal{F}_1, L_1(P_0)\}<\infty$.

With respect to the envelop function, 
\begin{align*}
\left| \log \frac{f_0(x_1)}{f(x_1)} \right| & \leq \log \max\left( \overline{\tau} \tau^{-1}_{0,1}, \tau_{0,1} \underline{\tau}^{-1} \right) + (\tau_{0,1}^{-2}+\underline{\tau}) x_1^2 + 2( \tau_{0,1}^{-2} k + \underline{\tau}^{-2}k' )|x_1| \\
& \ \  + \tau_{0,1}^{-2}k^2 + \underline{\tau}^{-2} k'^2, \\ 
& \equiv B(x_1).
\end{align*}
It is easy to check $\int B(x_1) dP_0 < \infty$.  As a result, $[ \log\{ f_0(x_1)/f(x_1) \}, f\in \mathcal{F} ]$ is $P_0$-Glivenko-Cantelli.

%%%%%%%%%%%%%%%%%%%%%%%%%%%%%%%
\section{Supplemental materials for simulation study}
%%%%%%%%%%%%%%%%%%%%%%%%%%%%%%%

%%%%%%%%%%%%%%%%%%%%%%%%%%%%%%%
\subsection{Convergence check}

\begin{figure}[htbp]
\figurebox{15pc}{40pc}{}[conv-case1.eps]
\caption{Sample paths (top) and autocorrelations (bottom) of $\zeta_j(f,P_n)$ with $j=1,\ldots,5$ for Case 1.}
\end{figure}

\begin{figure}[htbp]
\figurebox{15pc}{40pc}{}[conv-case1-2.eps]
\caption{Sample paths (top) and autocorrelations (bottom) of $\zeta_j(f,P_n)$ with $j=6,\ldots,10$ for Case 1.}
\end{figure}

\begin{figure}[htbp]
\figurebox{15pc}{40pc}{}[conv-case2.eps]
\caption{Sample paths (top) and autocorrelations (bottom) of $\zeta_j(f,P_n)$ with $j=1,\ldots,5$ for Case 2.}
\end{figure}

\begin{figure}[htbp]
\figurebox{15pc}{40pc}{}[conv-case2-2.eps]
\caption{Sample paths (top) and autocorrelations (bottom) of $\zeta_j(f,P_n)$ with $j=6,\ldots,10$ for Case 2.}
\end{figure}

\begin{figure}[htbp]
\figurebox{15pc}{40pc}{}[conv-case3.eps]
\caption{Sample paths (top) and autocorrelations (bottom) of $\zeta_j(f,P_n)$ with $j=1,\ldots,5$ for Case 3.}
\end{figure}

\begin{figure}[htbp]
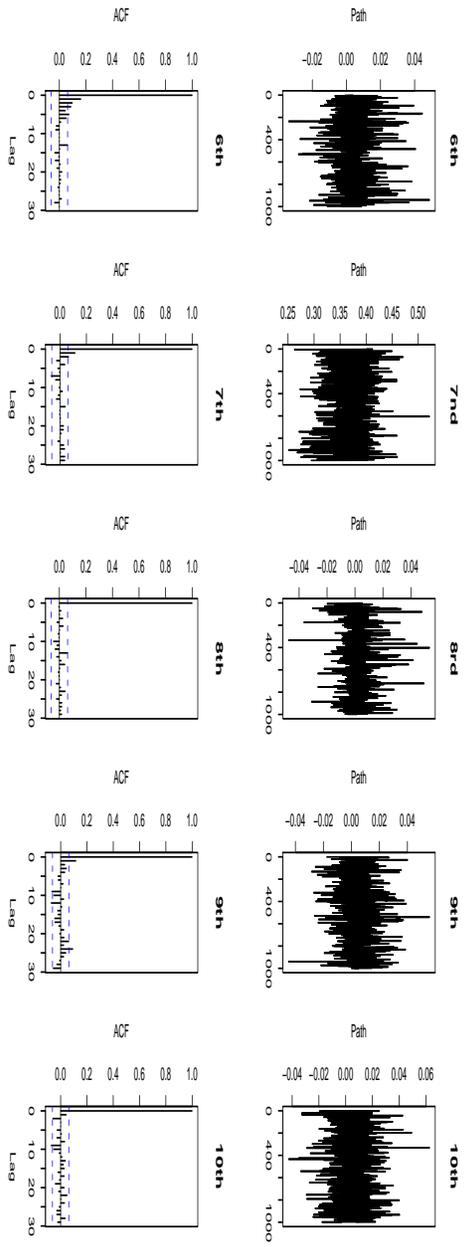

\figurebox{15pc}{40pc}{}[conv-case3-2.eps]
\caption{Sample paths (top) and autocorrelations (bottom) of $\zeta_j(f,P_n)$ with $j=6,\ldots,10$ for Case 3.}
\end{figure}

\clearpage

%%%%%%%%%%%%%%%%%%%%%%%%%%%%%%%
\subsection{Detecting marginal relationships}

To assess type I error rates, we applied two examples of null distributions in \cite{Heller-et-al13} with $n=100$. The first one is named four independent clouds for which we generated two univariate variables $y_i$ and $x_i$ identically and independently from $0.5N(-1,0.2)+0.5N(1,0.2)$ for $i=1,\ldots,n$.  As a competitor, we use R package implementation of the \cite{Heller-et-al13} method with default settings using 1,000 random permutations and 0.05 significance level.  Also, we use the same Markov chain Monte Carlo settings as in the simulation study for our proposed method.  The type 1 error rates of the proposed method and the competitors over 100 data sets are 0.05 and 0.04 respectively.  In the second example, all variables are identically and independently distributed from $N(0,1)$ with a univariate $y_i$ and $x_i=(x_{i,1},\ldots,x_{i,p})^{T}$ with $p=10$.  The type 1 error rates are 0.00 and 0.02 for the proposed method and the competitor.

\begin{figure}[htp]
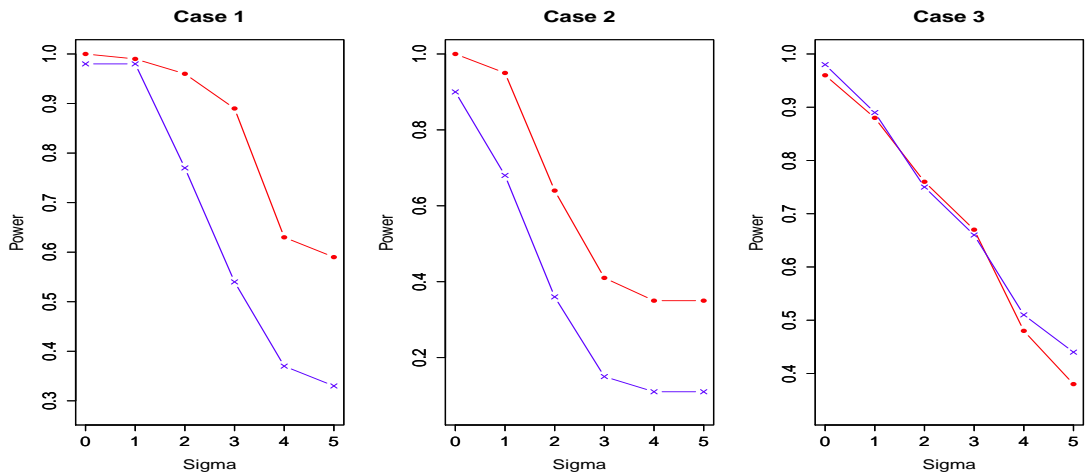

\figurebox{15.5pc}{35pc}{}[Power.eps]
\caption{Comparison of power by the proposed method (red) and \cite{Heller-et-al13} (blue) in Case 1 (left), Case 2 (middle) and Case 3 (right). $y$-axis indicates the power over 100 simulations and $x$-axis shows the standard deviation of the additional error term.}
\label{hhg}
\end{figure}

With respect to power, we first generate $y_i$ and $x_i=(x_{i,1},\ldots,x_{i,p})^{\T}$ in each of Case 1, 2 and 3 and put an additional error, $y_i^*=y_i+\varepsilon_i^*$ where $\{\varepsilon_i^*\}$ are independent and identically distributed from $N(0,\sigma^{*2})$.  Then, we checked the performance of detecting dependence between $y_i^*$ and $x_i$ with $\sigma^*=0,1,2,3,4,5$.  Figure \ref{hhg} reports the power estimated from 100 data sets in each case.  Although Case 3 shows little difference between the two methods, the proposed method outperforms \cite{Heller-et-al13} with relatively large difference in Case 1 and 2.

%%%%%%%%%%%%%%%%%%%%%%%%%%%%%%%
\section{Supplemental materials for application to criminology data}
%%%%%%%%%%%%%%%%%%%%%%%%%%%%%%%

%%%%%%%%%%%%%%%%%%%%%%%%%%%%%%%
\subsection{Data in the criminology application}
%%%%%%%%%%%%%%%%%%%%%%%%%%%%%%%

The whole data set can be downloaded from the University of California Irvine machine learning repository website.  Further information is given in [1] United States Department of Commerce, Bureau of the Census, census of population and housing 1990 United States: summary tape file 1a and 3a, [2] United States Department of Commerce, Bureau of the Census Producer, Washington, DC and Inter-university consortium for political and social research, Ann Arbor, Michigan in 1992, [3] United States Department of Justice, Bureau of Justice Statistics, law enforcement management and administrative statistics, [4] United States Department of Justice, Federal Bureau of Investigation, crime in the United States in 1995.

As for the predictors, Table \ref{tb:list1} and Table \ref{tb:list2} give the whole list.

%%%%%%%%%%%%%%%%%%%%%%%%%%%%%%%%%
\begin{table}[h]
\tbl{List of 1st to 34th predictors}{%
\begin{tabular}{cll}
\hline
\multicolumn{1}{c}{No.}		& \multicolumn{1}{l}{Predictor}	&  \multicolumn{1}{l}{Scale} \\
\hline 
1 & population for community &  count \\
2 & mean people per household & continuous \\
3 & \% of population that is african american & percent \\ 
4 & \% of population that is caucasian & percent \\ 
5 & \% of population that is of asian heritage & percent \\ 
6 & \% of population that is of hispanic heritage & percent \\ 
7 & \% of population that is 16-24 in age & percent \\ 
8 & \% of population that is 65 and over in age & percent \\ 
9 & \# of people living in areas classified as urban & count \\ 
10 & median household income & continuous \\
11 & \% of households with wage or salary income in 1989 & percent \\ 
12 & \% of households with farm or self employment income in 1989 & percent \\ 
13 & \% of households with investment / rent income in 1989 & percent\\ 
14 & \% of households with social security income in 1989 & percent \\ 
15 & \% of households with public assistance income in 1989 & percent \\ 
16 & \% of households with retirement income in 1989 & percent \\ 
17 & median family income & continuous \\ 
18 & per capita income & continuous \\ 
19 & \# of people under the poverty level & count \\ 
20 & \% of people 25 and over with less than a 9th grade education & percent \\ 
21 & \% of people 25 and over that are not high school graduates & percent \\ 
22 & \% of people 25 and over with a bachelors degree or higher education & percent \\ 
23 & \% of people 16 and over, in the labor force, and unemployed & percent \\
24 & \% of people 16 and over who are employed & percent \\ 
25 & \% of people 16 and over who are employed in manufacturing & percent \\ 
26 & \% of people 16 and over who are employed in professional services & percent \\ 
27 & \% of males who are divorced & percent \\
28 & \% of males who have never married & percent\\ 
29 & \% of females who are divorced & percent\\ 
30 & \% of population who are divorced & percent \\ 
31 & mean number of people per family & continuous \\ 
32 & \% of families (with kids) that are headed by two parents & percent \\ 
33 & \% of kids in family housing with two parents & percent \\ 
34 & \% of kids 4 and under in two parent households & percent \\ 
\hline
\end{tabular}}
\label{tb:list1}
\end{table}
%%%%%%%%%%%%%%%%%%%%%%%%%%%%%%%%%%%%%%%%%%%%%%%%%%%%%%%%%%%%%%%%%%%%%%%%%%%%%%%%%%%%%%%%%%%%

%%%%%%%%%%%%%%%%%%%%%%%%%%%%%%%%%
\begin{table}[h]
\tbl{List of 35th to 68th predictors}{%
\begin{tabular}{cll}
\hline
\multicolumn{1}{c}{No.}		& \multicolumn{1}{l}{Predictor} &  \multicolumn{1}{l}{Scale}	\\
\hline 
35 & \% of kids age 12-17 in two parent households & percent \\ 
36 & \% of moms of kids 6 and under in labor force & percent \\ 
37 & \% of moms of kids under 18 in labor force & percent \\ 
38 & \# of kids born to never married & count \\ 
39 & total number of people known to be foreign born & count \\ 
40 & \% of immigrants who immigated within last 5 years & percent \\ 
41 & \% of population who have immigrated within the last 5 years & percent \\ 
42 & \% of people who speak only English & percent \\ 
43 & \% of people who do not speak English well & percent \\ 
44 & \% of family households that are large (6 or more) & percent \\ 
45 & \% of all occupied households that are large (6 or more people) & percent \\
46 & \% of people in owner occupied households & percent \\ 
47 & \% of persons in dense housing (more than 1 person per room) & percent \\ 
48 & \% of housing units with less than 3 bedrooms & percent \\
49 & \# of vacant households & count \\
50 & \% of housing occupied & percent \\ 
51 & \% of households owner occupied & percent \\ 
52 & \% of vacant housing that is boarded up & percent \\ 
53 & \% of vacant housing that has been vacant more than 6 months & percent \\ 
54 & owner occupied housing: lower quartile value & continuous \\
55 & owner occupied housing: median value & continuous \\ 
56 & owner occupied housing: upper quartile value & continuous \\ 
57 & rental housing: lower quartile rent & continuous \\
58 & rental housing: median rent & continuous \\
59 & rental housing: upper quartile rent & continuous \\ 
60 & median gross rent & continuous \\
61 & median gross rent as \% of household income & percent \\ 
62 & \# of people in homeless shelters & count \\
63 & \# of homeless people counted in the street & count \\ 
64 & \% of people born in the same state as currently living & percent \\
65 & \% of people living in the same city as in 1985 (5 years before) & percent \\ 
66 & land area in square miles & continuous \\
67 & population density in persons per square mile & continuous \\ 
68 & \% of people using public transit for commuting & percent \\   
\hline
\end{tabular}}
\label{tb:list2}
\end{table}
%%%%%%%%%%%%%%%%%%%%%%%%%%%%%%%%%%%%%%%%%%%%%%%%%%%%%%%%%%%%%%%%%%%%%%%%%%%%%%%%%%%%%%%%%%%%

%%%%%%%%%%%%%%%%%%%%%
\subsection{Markov chain Monte Carlo Algorithm} \label{ap:mcmc}
%%%%%%%%%%%%%%%%%%%%%

Relying on the blocked Gibbs sampler by \cite{IshwaranJames01}, we develop an efficient posterior computation method for the Dirichlet process mixture model in Section 4.  Let $s=(s_1,\ldots,s_n)'$ be the latent cluster index variables.   Then, we propose the following Markov chain Monte Carlo algorithm:

\begin{step}
Update $V_h$ for $h=1,\ldots,H-1$ from
\begin{align*}
\text{Be} \left( 1+n_h, \alpha_0+\sum_{l> h} n_l \right),
\end{align*}
where $n_h=\sum_{i=1}^n 1(s_i=h)$.
\end{step}

\begin{step}
Using the prior $\text{Gamma}(a_{\alpha}, b_{\alpha})$, update $\alpha_0$ from
\begin{align*}
\text{Gamma} \left\{ a_{\alpha}+H-1, b_{\alpha} - \sum_{h=1}^{H-1} \log(1-V_h) \right\}.
\end{align*}
\end{step}

\begin{step}
Update $s_{i}$ for $i=1,\ldots,n$ from  
\begin{align*}
\text{pr}(s_{i}=h \mid \cdots) &= \frac{ \pi_h  f(y_i \mid x_i, \theta_h) \prod_{j=1}^p f(x_{i,j} \mid \theta_h) }{ \sum_{l=1}^{H} \pi_l f(y_i \mid x_i, \theta_l) \prod_{j=1}^p f(x_{i,j} \mid \theta_l) }. 
\end{align*}
\end{step}

\begin{step}
Update $\mu_{j,h}$ for $j=1,\ldots,p$ and $h=1,\ldots,H$ from $N(\tilde{\mu}_{j,h},\tilde{\tau}^2_{j,h})$ where 
\begin{align*}
\tilde{\mu}_{j,h} = \tilde{\tau}^2_{j,h} \left( \frac{\sum_{i:s_{i}=h} x_{i,j}}{\tau^2_{j,h}} + \frac{\bar{\mu}_{j}}{s_j^2} \right), \ \ \tilde{\tau}^2_{j,h} = \left(\frac{n_{h}}{\tau^2_{j,h}} + \frac{1}{s_j^2} \right)^{-1}, \ \ n_{h} = \sum_{i=1}^n 1(s_{i}=h).
\end{align*}
\end{step}

\begin{step}
Update $\tau^2_{j,h}$ for $j=1,\ldots,p$ and $h=1,\ldots,H$ from 
\begin{align*}
\text{IG} \left\{ \frac{n_{h}+3}{2}, \frac{ \sum_{i:s_{i=h}} (x_{i,j}-\mu_{j,h})^2 + s_j^2}{2} \right\}.
\end{align*}
\end{step}

\begin{step}
Update $\sigma^2_{h}$ for $h=1,\ldots,H$ from 
\begin{align*}
\text{IG} \left\{ \frac{n_h+3}{2}, \frac{\sum_{i:s_{i}=h} (y_{i}-\tilde{x}^{\T}_i \beta_h )^2 + s_y^2}{2} \right\}.
\end{align*}
\end{step}

\begin{step}
Update $\beta_{j,h}$ for $j=0,\ldots,p$ and $h=1,\ldots,H$ from
\begin{align*}
\pi(\beta_{j,h}\mid \cdots) = \hat{p}_{j,h} \delta_{0}(\beta_{j,h}) + (1-\hat{p}_{j,h}) N(\beta_{j,h}\mid \mu_{\beta_{j,h}}, \sigma^2_{\beta_{j,h}}), \\
\end{align*}
where
\begin{align*}
\mu_{\beta_{j,h}} &= \sigma^2_{\beta_{j,h}} \left\{ \sum_{i:s_{i}=h} \frac{x_{i,j}\left( y_i-\tilde{x}^{\T}_{i,-j} \beta_{-j,h} \right)}{\sigma^2_{h}} \right\}, \ \ \sigma^2_{\beta_{j,h}} = \left( \sum_{i:s_{i}=h} \frac{x_{i,j}^2}{\sigma^2_{h}} + \frac{1}{\lambda_{j,h}^2} \right)^{-1}, \\
\hat{p}_{j,h} &= \left\{ 1 +  \frac{1-p_{0j}}{p_{0j}} \frac{N(0\mid 0, \lambda_{j,h}^2)}{N(0\mid \mu_{\beta_{j,h}}, \sigma^2_{\beta_{j,h}})} \right\}^{-1}.
\end{align*}
\end{step}

\begin{step}
Update $\lambda_{j,h}^2$ for $j=1,\ldots,p$ and $h=1,\ldots,H$ from
\begin{align*}
\text{IG} \left\{ \frac{ 1(\beta_{j,h}\neq 0) + 1}{2}, \frac{ \beta_{j,h}^2 + 1}{2} \right\}.
\end{align*}
\end{step}

\begin{step}
Update $p_0$ from
\begin{align*}
\text{Be} \left\{ 4.75 + \sum_{j,h} 1(\beta_{j,h}= 0), 0.25 + \sum_{j,h} 1(\beta_{j,h}\neq 0) \right\}.
\end{align*}
\end{step}

\begin{step}
Impute missing values $y^{\text{mis}}_i$ in the response.
\begin{algorithm}
\vspace*{-12pt}
\begin{tabbing}
   \enspace 1. Generate $y^*_i \sim N(\tilde{x}^{\T}_i \beta_{s_i}, \sigma^2_{s_i})$.\\
   \enspace 2. Set $y_i^{\text{mis}} = l$ if $a_l < y_i^* \leq a_{l+1}$.
\end{tabbing}
\end{algorithm}
\end{step}

\begin{step}
Update latent variables $y^*_i$ and $x^*_{i,j}$ for count and percentage variables.
\begin{algorithm}
\vspace*{-12pt}
\begin{tabbing}
   \enspace (a) For the response variable, $y_i^* \sim TN(\tilde{x}^{\T}_i \beta_{s_i}, \sigma^2_{s_i}, a_{y_i}, a_{y_{i+1}})$, \\
   \enspace (b) For the count predictor, $x_{i,j}^* \sim TN(\mu_{j,s_i}, \tau^2_{j,s_i}, a_{x_i}, a_{x_{i+1}})$, \\
   \enspace (c) For the percentage predictor, $x_{i,j}^* \sim TN(\mu_{j,s_i}, \tau^2_{j,s_i}, -\infty, 0)$ if $x_{i,j}=0$ \\ 
   \enspace and $x_{i,j}^* \sim TN(\mu_{j,s_i}, \tau^2_{j,s_i}, 100, \infty)$ if $x_{i,j}=100$, 
\end{tabbing}
\end{algorithm}

where $TN(a,b,c,d)$ denotes a truncated normal with the location $a$, scale $b$, lower bound $c$ and upper bound $d$.
\end{step}

\begin{step}
Compute and save $\zeta_{j}(f,P_n)$ for $j=1,\ldots,p$.
\end{step}

%%%%%%%%%%%%%%%%%%%%%%%%%%%%%%%
\subsection{Convergence check}

\begin{figure}[htbp]
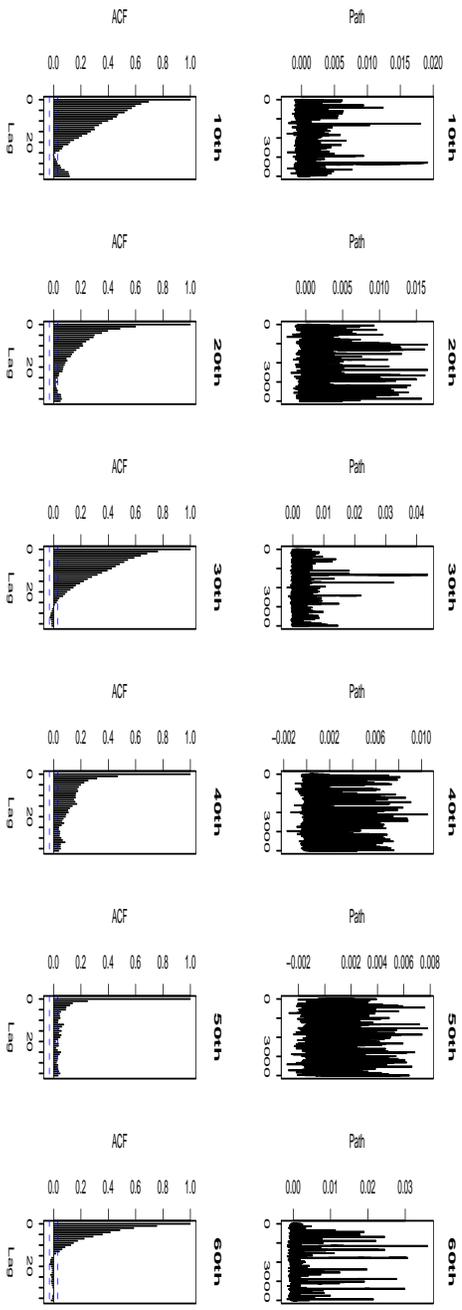

\figurebox{15pc}{42.5pc}{}[conv-murder2.eps]
\caption{Sample paths (top) and autocorrelations (bottom) of $\zeta_j(f,P_n)$ with $j=10,20,30,40,50,60$ for murders.}
\label{conv-murder}
\end{figure}

\begin{figure}[htbp]
\figurebox{15pc}{42.5pc}{}[conv-rape2.eps]
\caption{Sample paths (top) and autocorrelations (bottom) of $\zeta_j(f,P_n)$ with $j=10,20,30,40,50,60$ for rapes.}
\label{conv-rape}
\end{figure}

\begin{figure}[htbp]
\figurebox{15pc}{42.5pc}{}[conv-robbery2.eps]
\caption{Sample paths (top) and autocorrelations (bottom) of $\zeta_j(f,P_n)$ with $j=10,20,30,40,50,60$ for robberies.}
\label{conv-robbery}
\end{figure}

\begin{figure}[htbp]
\figurebox{15pc}{42.5pc}{}[conv-assault2.eps]
\caption{Sample paths (top) and autocorrelations (bottom) of $\zeta_j(f,P_n)$ with $j=10,20,30,40,50,60$ for assaults.}
\label{conv-assault}
\end{figure}

\begin{figure}[htbp]
\figurebox{15pc}{42.5pc}{}[conv-burglary2.eps]
\caption{Sample paths (top) and autocorrelations (bottom) of $\zeta_j(f,P_n)$ with $j=10,20,30,40,50,60$ for burglaries.}
\label{conv-burglary}
\end{figure}

\begin{figure}[htbp]
\figurebox{15pc}{42.5pc}{}[conv-larceny2.eps]
\caption{Sample paths (top) and autocorrelations (bottom) of $\zeta_j(f,P_n)$ with $j=10,20,30,40,50,60$ for larcenies.}
\label{conv-larceny}
\end{figure}

\begin{figure}[htbp]
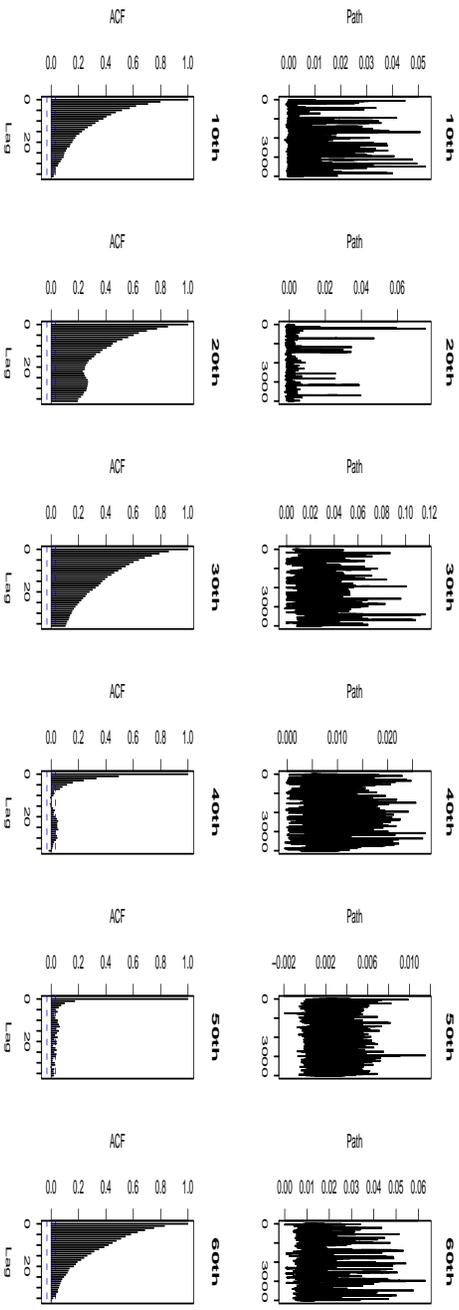

\figurebox{15pc}{42.5pc}{}[conv-autotheft2.eps]
\caption{Sample paths (top) and autocorrelations (bottom) of $\zeta_j(f,P_n)$ with $j=10,20,30,40,50,60$ for auto thefts.}
\label{conv-autotheft}
\end{figure}

\begin{figure}[htbp]
\figurebox{15pc}{42.5pc}{}[conv-arson2.eps]
\caption{Sample paths (top) and autocorrelations (bottom) of $\zeta_j(f,P_n)$ with $j=10,20,30,40,50,60$ for arsons.}
\label{conv-arson}
\end{figure}

\begin{figure}[htbp]
\figurebox{15pc}{42.5pc}{}[conv-violent2.eps]
\caption{Sample paths (top) and autocorrelations (bottom) of $\zeta_j(f,P_n)$ with $j=10,20,30,40,50,60$ for violent crimes.}
\label{conv-violent}
\end{figure}

\begin{figure}[htbp]
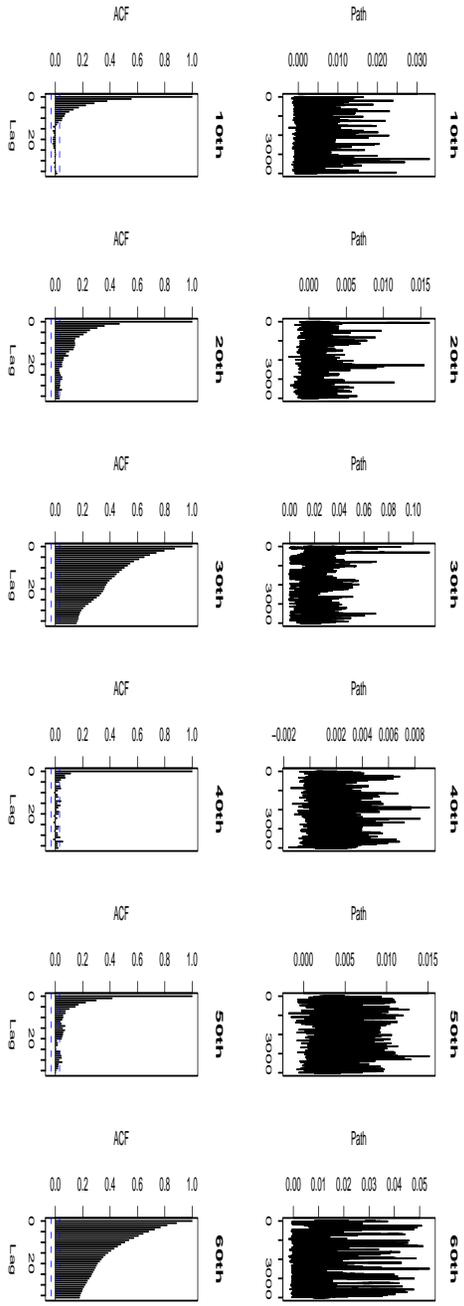

\figurebox{15pc}{42.5pc}{}[conv-nonviolent2.eps]
\caption{Sample paths (top) and autocorrelations (bottom) of $\zeta_j(f,P_n)$ with $j=10,20,30,40,50,60$ for non violent crimes.}
\label{conv-nonviolent}
\end{figure}

\clearpage

%%%%%%%%%%%%%%%%%%%%%%%%%%%%%%%
\subsection{Additional estimation results}
%%%%%%%%%%%%%%%%%%%%%%%%%%%%%%%

Tables \ref{tb:list3}-\ref{tb:list12} show lists of the selected predictors by the proposed method for murders, rapes, robberies, assaults, burglaries, larcenies, auto thefts, arsons, violent crimes and non-violent crimes, respectively.  The predictors are listed in descending order of the posterior mean of the conditional mutual information.  Also, 90\% credible intervals of the conditional mutual information are reported in Figure \ref{fig1}-\ref{fig9} for all crime variables except murders.

Tables \ref{tb:list13} and \ref{tb:list14} report lists of the selected predictors by the competitors.  Results for murders, rapes, robberies, assaults, burglaries and larcenies are in Table \ref{tb:list13} and those for auto thefts, arsons, violent crimes and non-violent crimes are in Table \ref{tb:list14}.

%%%%%%%%%%%%%%%%%%%%%%%%%%%%%%%
\subsection{Discussion of alternative approach}
%%%%%%%%%%%%%%%%%%%%%%%%%%%%%%%

One possible approach of measuring conditional independence may be to estimate conditional mutual information based on the empirical measure and the kernel density estimation of the joint density instead of the nonparametric Bayes encompassing model.  However, \cite{Joe89} and \cite{SethPrincipe12b} point out high sensitivity of the estimation result depending on the choice of the kernel and its band-width.  Especially in a case with not a small $p$, it may not straightforward to choose them appropriately.  Therefore, the key is to develop a kernel method which produces a stable estimation result.

%%%%%%%%%%%%%%%%%%%%%%%%%%%%%%%%%
\begin{table}[htp]
\tbl{List of the selected predictors by the proposed method in descending order of the posterior means of conditional mutual information with murders as the response}{%
\begin{tabular}{cccl}
\hline
\multicolumn{1}{c}{$j$} & \multicolumn{1}{c}{Mean}	& \multicolumn{1}{c}{90\%CI}	& \multicolumn{1}{l}{Predictor}	\\
\hline 
66 & 0.2587 & [0.2157, 0.2936] & land area in square miles \\
67 & 0.1188 & [0.0905, 0.1454] & population density in persons per square mile \\ 
4  & 0.0507 & [0.0302, 0.0678] & \% of population that is caucasian  \\ 
9  & 0.0250 & [0.0043, 0.0636] & \# of people living in areas classified as urban \\ 
1  & 0.0250 & [0.0015, 0.0469] & population for community  \\ 
3  & 0.0192 & [0.0058, 0.0374] & \% of population that is african american \\ 
57 & 0.0177 & [0.00007, 0.0463] & rental housing: lower quartile rent \\ 
13 & 0.0075 & [0.0004, 0.0149] & \% of households with investment / rent income in 1989 \\ 
6  & 0.0067 & [0.0021, 0.0125] & \% of population that is of hispanic heritage  \\ 
64 & 0.0039 & [0.0005, 0.0067] & \% of people born in the same state as currently living  \\ 
49 & 0.0030 & [0.0003, 0.0089] & \# of vacant households  \\ 
42 & 0.0027 & [0.0001, 0.0092] & \% of people who speak only English \\ 
27 & 0.0019 & [0.0001, 0.0055] & \% of males who are divorced \\ 
52 & 0.0018 & [0.0002, 0.0051] & \% of vacant housing that is boarded up \\ 
\hline
\end{tabular}}
\begin{tabnote}
$j$, $j$-th predictor; Mean, posterior mean; 90\%CI refers to a 90\% credible interval.
\end{tabnote}
\label{tb:list3}
\end{table}
%%%%%%%%%%%%%%%%%%%%%%%%%%%%%%%%%%%%%%%%%%%%%%%%%%%%%%%%%%%%%%%%%%%%%%%%%%%%%%%%%%%%%%%%%%%%

%%%%%%%%%%%%%%%%%%%%%%%%%%%%%%%%%
\begin{table}[htp]
\tbl{List of the selected predictors by the proposed method  in descending order of the posterior means of conditional mutual information with rapes as the response}{%
\begin{tabular}{cccl}
\hline
\multicolumn{1}{c}{$j$} & \multicolumn{1}{c}{Mean}	& \multicolumn{1}{c}{90\%CI}	& \multicolumn{1}{l}{Predictor}	\\
\hline 
66 & 0.4168 & [0.3929, 0.4428] & land area in square miles \\
67 & 0.1964 & [0.1727, 0.2217] & population density in persons per square mile \\ 
1  & 0.0680 & [0.0523, 0.0865] & population for community  \\ 
9  & 0.0359 & [0.0086, 0.0608] & \# of people living in areas classified as urban \\ 
30 & 0.0189 & [0.0013, 0.0379] & \% of population who are divorced \\ 
32 & 0.0178 & [0.0009, 0.0398] & \% of families (with kids) that are headed by two parents \\ 
33 & 0.0174 & [0.0006, 0.0389] & \% of kids in family housing with two parents  \\ 
29 & 0.0156 & [0.0005, 0.0330] & \% of females who are divorced  \\ 
27 & 0.0123 & [0.0009, 0.0265] & \% of males who are divorced \\ 
39 & 0.0051 & [0.0004, 0.0118] & total number of people known to be foreign born  \\
5  & 0.0046 & [0.0002, 0.0092] & \% of population that is of asian heritage  \\ 
35 & 0.0031 & [0.0001, 0.0125] & \% of kids age 12-17 in two parent households \\ 
7  & 0.0027 & [0.0008, 0.0056] & \% of population that is 16-24 in age  \\ 
50 & 0.0023 & [0.0002, 0.0053] & \% of housing occupied \\ 
12 & 0.0022 & [0.0001, 0.0059] & \% of households with farm or self employment income in 1989 \\ 
19 & 0.0021 & [0.0003, 0.0062] & \# of people under the poverty level  \\ 
38 & 0.0017 & [0.0001, 0.0065] & \# of kids born to never married  \\ 
18 & 0.0015 & [0.00008, 0.0050] & per capita income \\ 
28 & 0.0014 & [0.0002, 0.0036] & \% of males who have never married \\ 
63 & 0.0011 & [0.0002, 0.0020] & \# of homeless people counted in the street \\ 
\hline
\end{tabular}}
\begin{tabnote}
$j$, $j$-th predictor; Mean, posterior mean; 90\%CI refers to a 90\% credible interval.
\end{tabnote}
\label{tb:list4}
\end{table}
%%%%%%%%%%%%%%%%%%%%%%%%%%%%%%%%%%%%%%%%%%%%%%%%%%%%%%%%%%%%%%%%%%%%%%%%%%%%%%%%%%%%%%%%%%%%

%%%%%%%%%%%%%%%%%%%%%%%%%%%%%%%%%
\begin{table}[htp]
\tbl{List of the selected predictors by the proposed method  in descending order of the posterior means of conditional mutual information with robberies as the response}{%
\begin{tabular}{cccl}
\hline
\multicolumn{1}{c}{$j$} & \multicolumn{1}{c}{Mean}	& \multicolumn{1}{c}{90\%CI}	& \multicolumn{1}{l}{Predictor}	\\
\hline 
66 & 0.6074 & [0.5551, 0.6554] & land area in square miles \\
67 & 0.5080 & [0.4548, 0.5605] & population density in persons per square mile \\ 
33 & 0.0859 & [0.0545, 0.1203] & \% of kids in family housing with two parents  \\ 
4  & 0.0652 & [0.0353, 0.0953] & \% of population that is caucasian  \\ 
3  & 0.0530 & [0.0211, 0.0865] & \% of population that is african american \\ 
9  & 0.0469 & [0.0078, 0.0926] & \# of people living in areas classified as urban \\ 
1  & 0.0388 & [0.0268, 0.0623] & population for community  \\ 
47 & 0.0277 & [0.0084, 0.0493] & \% of persons in dense housing (more than 1 person per room) \\ 
30 & 0.0159 & [0.0007, 0.0348] & \% of population who are divorced \\ 
18 & 0.0139 & [0.0009, 0.0326] & per capita income \\
32 & 0.0122 & [0.0006, 0.0340] & \% of families (with kids) that are headed by two parents \\ 
29 & 0.0107 & [0.0002, 0.0258] & \% of females who are divorced  \\ 
6  & 0.0106 & [0.0006, 0.0237] & \% of population that is of hispanic heritage  \\ 
64 & 0.0094 & [0.0045, 0.0146] & \% of people born in the same state as currently living  \\ 
42 & 0.0090 & [0.0002, 0.0217] & \% of people who speak only English \\ 
22 & 0.0079 & [0.0001, 0.0198] & \% of people 25 and over with a bachelors degree or higher education \\ 
46 & 0.0071 & [0.0006, 0.0183] & \% of people in owner occupied households \\ 
56 & 0.0064 & [0.0001, 0.0182] & owner occupied housing: upper quartile value \\ 
25 & 0.0062 & [0.0002, 0.0125] & \% of people 16 and over who are employed in manufacturing \\ 
68 & 0.0055 & [0.0015, 0.0099] & \% of people using public transit for commuting \\ 
34 & 0.0054 & [0.0004, 0.0183] & \% of kids 4 and under in two parent households \\ 
51 & 0.0050 & [0.0006, 0.0142] & \% of households owner occupied \\ 
19 & 0.0030 & [0.0003, 0.0072] & \# of people under the poverty level  \\ 
38 & 0.0029 & [0.0005, 0.0077] & \# of kids born to never married  \\ 
49 & 0.0021 & [0.0001, 0.0056] & \# of vacant households  \\ 
\hline
\end{tabular}}
\begin{tabnote}
$j$, $j$-th predictor; Mean, posterior mean; 90\%CI refers to a 90\% credible interval.
\end{tabnote}
\label{tb:list5}
\end{table}
%%%%%%%%%%%%%%%%%%%%%%%%%%%%%%%%%%%%%%%%%%%%%%%%%%%%%%%%%%%%%%%%%%%%%%%%%%%%%%%%%%%%%%%%%%%%

%%%%%%%%%%%%%%%%%%%%%%%%%%%%%%%%%
\begin{table}[htp]
\tbl{List of the selected predictors by the proposed method  in descending order of the posterior means of conditional mutual information with assaults as the response}{%
\begin{tabular}{cccl}
\hline
\multicolumn{1}{c}{$j$} & \multicolumn{1}{c}{Mean}	& \multicolumn{1}{c}{90\%CI}	& \multicolumn{1}{l}{Predictor}	\\
\hline 
66 & 0.3380 & [0.2897, 0.3914] & land area in square miles \\
67 & 0.1760 & [0.1318, 0.2267] & population density in persons per square mile \\ 
9  & 0.0760 & [0.0451, 0.0996] & \# of people living in areas classified as urban \\ 
1  & 0.0413 & [0.0186, 0.0641] & population for community  \\ 
33 & 0.0350 & [0.0114, 0.0571] & \% of kids in family housing with two parents  \\ 
13 & 0.0348 & [0.0234, 0.0478] & \% of households with investment / rent income in 1989 \\ 
32 & 0.0176 & [0.0010, 0.0403] & \% of families (with kids) that are headed by two parents \\ 
47 & 0.0171 & [0.0057, 0.0283] & \% of persons in dense housing (more than 1 person per room) \\ 
4  & 0.0168 & [0.0046, 0.0284] & \% of population that is caucasian  \\ 
3  & 0.0070 & [0.0004, 0.0174] & \% of population that is african american \\ 
43 & 0.0050 & [0.0013, 0.0102] & \% of people who do not speak English well \\ 
45 & 0.0027 & [0.0003, 0.0074] & \% of all occupied households that are large (6 or more people) \\ 
50 & 0.0025 & [0.0007, 0.0046] & \% of housing occupied \\ 
34 & 0.0024 & [0.0001, 0.0075] & \% of kids 4 and under in two parent households \\ 
44 & 0.0023 & [0.0003, 0.0064] & \% of family households that are large (6 or more) \\ 
23 & 0.0014 & [0.0001, 0.0041] & \% of people 16 and over, in the labor force, and unemployed \\ 
\hline
\end{tabular}}
\begin{tabnote}
$j$, $j$-th predictor; Mean, posterior mean; 90\%CI refers to a 90\% credible interval.
\end{tabnote}
\label{tb:list6}
\end{table}
%%%%%%%%%%%%%%%%%%%%%%%%%%%%%%%%%%%%%%%%%%%%%%%%%%%%%%%%%%%%%%%%%%%%%%%%%%%%%%%%%%%%%%%%%%%%

%%%%%%%%%%%%%%%%%%%%%%%%%%%%%%%%%
\begin{table}[htp]
\tbl{List of the selected predictors by the proposed method  in descending order of the posterior means of conditional mutual information with burglaries as the response}{%
\begin{tabular}{cccl}
\hline
\multicolumn{1}{c}{$j$} & \multicolumn{1}{c}{Mean}	& \multicolumn{1}{c}{90\%CI}	& \multicolumn{1}{l}{Predictor}	\\
\hline 
66 & 0.9177 & [0.8717, 0.9492] & land area in square miles \\
67 & 0.7075 & [0.6639, 0.7464] & population density in persons per square mile \\ 
33 & 0.0508 & [0.0241, 0.0796] & \% of kids in family housing with two parents  \\ 
47 & 0.0281 & [0.0146, 0.0444] & \% of persons in dense housing (more than 1 person per room) \\ 
29 & 0.0173 & [0.0100, 0.0276] & \% of females who are divorced  \\ 
50 & 0.0152 & [0.0071, 0.0236] & \% of housing occupied \\ 
13 & 0.0135 & [0.0008, 0.0303] & \% of households with investment / rent income in 1989 \\ 
6  & 0.0097 & [0.00007, 0.0166] & \% of population that is of hispanic heritage  \\ 
30 & 0.0083 & [0.0001, 0.0224] & \% of population who are divorced \\ 
9  & 0.0078 & [0.0004, 0.0258] & \# of people living in areas classified as urban \\ 
4  & 0.0070 & [0.0007, 0.0163] & \% of population that is caucasian  \\ 
68 & 0.0057 & [0.0004, 0.0126] & \% of people using public transit for commuting \\ 
65 & 0.0048 & [0.0001, 0.0116] & \% of people living in the same city as in 1985 (5 years before) \\ 
49 & 0.0046 & [0.0005, 0.0125] & \# of vacant households  \\ 
7  & 0.0031 & [0.0002, 0.0066] & \% of population that is 16-24 in age  \\ 
19 & 0.0028 & [0.0001, 0.0110] & \# of people under the poverty level  \\ 
61 & 0.0024 & [0.00008, 0.0069] & median gross rent as \% of household income  \\ 
36 & 0.0008 & [0.00001, 0.0025] & \% of moms of kids 6 and under in labor force  \\ 
\hline
\end{tabular}}
\begin{tabnote}
$j$, $j$-th predictor; Mean, posterior mean; 90\%CI refers to a 90\% credible interval.
\end{tabnote}
\label{tb:list7}
\end{table}
%%%%%%%%%%%%%%%%%%%%%%%%%%%%%%%%%%%%%%%%%%%%%%%%%%%%%%%%%%%%%%%%%%%%%%%%%%%%%%%%%%%%%%%%%%%%

%%%%%%%%%%%%%%%%%%%%%%%%%%%%%%%%%
\begin{table}[htp]
\tbl{List of the selected predictors by the proposed method  in descending order of the posterior means of conditional mutual information with larcenies as the response}{%
\begin{tabular}{cccl}
\hline
\multicolumn{1}{c}{$j$} & \multicolumn{1}{c}{Mean}	& \multicolumn{1}{c}{90\%CI}	& \multicolumn{1}{l}{Predictor}	\\
\hline 
66 & 0.9425 & [0.9149, 0.9682] & land area in square miles \\
67 & 0.8035 & [0.7707, 0.8359] & population density in persons per square mile \\ 
32 & 0.0305 & [0.0003, 0.0505] & \% of families (with kids) that are headed by two parents \\ 
2  & 0.0233 & [0.0126, 0.0397] & mean people per household \\ 
22 & 0.0219 & [0.00001, 0.0436] & \% of people 25 and over with a bachelors degree or higher education \\ 
35 & 0.0217 & [0.0085, 0.0383] & \% of kids age 12-17 in two parent households \\ 
65 & 0.0165 & [0.0062, 0.0256] & \% of people living in the same city as in 1985 (5 years before) \\ 
8  & 0.0163 & [0.0008, 0.0321] & \% of population that is 65 and over in age \\ 
45 & 0.0135 & [0.00002, 0.0520] & \% of all occupied households that are large (6 or more people) \\ 
33 & 0.0133 & [0.00002, 0.0422] & \% of kids in family housing with two parents  \\ 
7  & 0.0106 & [0.0002, 0.0180] & \% of population that is 16-24 in age  \\ 
68 & 0.0105 & [0.0062, 0.0154] & \% of people using public transit for commuting \\ 
25 & 0.0084 & [0.0056, 0.0111] & \% of people 16 and over who are employed in manufacturing \\ 
47 & 0.0070 & [0.0001, 0.0178] & \% of persons in dense housing (more than 1 person per room) \\ 
23 & 0.0054 & [0.0004, 0.0103] & \% of people 16 and over, in the labor force, and unemployed \\ 
42 & 0.0054 & [0.0003, 0.0163] & \% of people who speak only English \\ 
64 & 0.0053 & [0.0005, 0.0123] & \% of people born in the same state as currently living  \\ 
5  & 0.0038 & [0.0001, 0.0077] & \% of population that is of asian heritage  \\ 
14 & 0.0022 & [0.0003, 0.0056] & \% of households with social security income in 1989 \\ 
\hline
\end{tabular}}
\begin{tabnote}
$j$, $j$-th predictor; Mean, posterior mean; 90\%CI refers to a 90\% credible interval.
\end{tabnote}
\label{tb:list8}
\end{table}
%%%%%%%%%%%%%%%%%%%%%%%%%%%%%%%%%%%%%%%%%%%%%%%%%%%%%%%%%%%%%%%%%%%%%%%%%%%%%%%%%%%%%%%%%%%%

%%%%%%%%%%%%%%%%%%%%%%%%%%%%%%%%%
\begin{table}[htp]
\tbl{List of the selected predictors by the proposed method  in descending order of the posterior means of conditional mutual information with auto thefts as the response}{%
\begin{tabular}{cccl}
\hline
\multicolumn{1}{c}{$j$} & \multicolumn{1}{c}{Mean}	& \multicolumn{1}{c}{90\%CI}	& \multicolumn{1}{l}{Predictor}	\\
\hline 
66 & 0.7650 & [0.7310, 0.8011] & land area in square miles \\
67 & 0.6471 & [0.6098, 0.6847] & population density in persons per square mile \\ 
47 & 0.0298 & [0.0164, 0.0437] & \% of persons in dense housing (more than 1 person per room) \\ 
30 & 0.0245 & [0.0008, 0.0541] & \% of population who are divorced \\ 
18 & 0.0229 & [0.0001, 0.0626] & per capita income \\ 
13 & 0.0211 & [0.0054, 0.0405] & \% of households with investment / rent income in 1989 \\ 
46 & 0.0197 & [0.00001, 0.0899] & \% of people in owner occupied households \\ 
60 & 0.0138 & [0.0054, 0.0342] & median gross rent  \\ 
53 & 0.0119 & [0.0050, 0.0178] & \% of vacant housing that has been vacant more than 6 months   \\ 
4  & 0.0095 & [0.0004, 0.0214] & \% of population that is caucasian  \\ 
42 & 0.0087 & [0.0014, 0.0190] & \% of people who speak only English \\ 
12 & 0.0081 & [0.0045, 0.0120] & \% of households with farm or self employment income in 1989 \\ 
2  & 0.0081 & [0.0004, 0.0234] & mean people per household \\ 
68 & 0.0075 & [0.0022, 0.0138] & \% of people using public transit for commuting \\ 
40 & 0.0071 & [0.0032, 0.0144] & \% of immigrants who immigated within last 5 years \\ 
43 & 0.0041 & [0.00005, 0.0123] & \% of people who do not speak English well \\ 
58 & 0.0034 & [0.0007, 0.0106] & rental housing: median rent \\ 
59 & 0.0030 & [0.0005, 0.0138] & rental housing: upper quartile rent \\ 
57 & 0.0022 & [0.0005, 0.0057] & rental housing: lower quartile rent \\ 
50 & 0.0021 & [0.0002, 0.0047] & \% of housing occupied \\ 
\hline
\end{tabular}}
\begin{tabnote}
$j$, $j$-th predictor; Mean, posterior mean; 90\%CI refers to a 90\% credible interval.
\end{tabnote}
\label{tb:list9}
\end{table}
%%%%%%%%%%%%%%%%%%%%%%%%%%%%%%%%%%%%%%%%%%%%%%%%%%%%%%%%%%%%%%%%%%%%%%%%%%%%%%%%%%%%%%%%%%%%

%%%%%%%%%%%%%%%%%%%%%%%%%%%%%%%%%
\begin{table}[htp]
\tbl{List of the selected predictors by the proposed method  in descending order of the posterior means of conditional mutual information with arsons as the response}{%
\begin{tabular}{cccl}
\hline
\multicolumn{1}{c}{$j$} & \multicolumn{1}{c}{Mean}	& \multicolumn{1}{c}{90\%CI}	& \multicolumn{1}{l}{Predictor}	\\
\hline 
66 & 0.3030 & [0.2517, 0.3593] & land area in square miles \\
67 & 0.1619 & [0.1226, 0.2084] & population density in persons per square mile \\ 
1  & 0.0394 & [0.0131, 0.0689] & population for community  \\ 
9  & 0.0152 & [0.0010, 0.0471] & \# of people living in areas classified as urban \\ 
19 & 0.0131 & [0.0005, 0.0323] & \# of people under the poverty level  \\ 
27 & 0.0119 & [0.0022, 0.0229] & \% of males who are divorced \\ 
13 & 0.0085 & [0.0004, 0.0168] & \% of households with investment / rent income in 1989 \\ 
29 & 0.0078 & [0.0001, 0.0212] & \% of females who are divorced  \\ 
41 & 0.0039 & [0.0013, 0.0071] & \% of population who have immigrated within the last 5 years \\ 
15 & 0.0031 & [0.0004, 0.0065] & \% of households with public assistance income in 1989  \\ 
\hline
\end{tabular}}
\begin{tabnote}
$j$, $j$-th predictor; Mean, posterior mean; 90\%CI refers to a 90\% credible interval.
\end{tabnote}
\label{tb:list10}
\end{table}
%%%%%%%%%%%%%%%%%%%%%%%%%%%%%%%%%%%%%%%%%%%%%%%%%%%%%%%%%%%%%%%%%%%%%%%%%%%%%%%%%%%%%%%%%%%%

%%%%%%%%%%%%%%%%%%%%%%%%%%%%%%%%%
\begin{table}[htp]
\tbl{List of the selected predictors by the proposed method  in descending order of the posterior means of conditional mutual information with violent crimes as the response}{%
\begin{tabular}{cccl}
\hline
\multicolumn{1}{c}{$j$} & \multicolumn{1}{c}{Mean}	& \multicolumn{1}{c}{90\%CI}	& \multicolumn{1}{l}{Predictor}	\\
\hline 
66 & 0.5254 & [0.4868, 0.5763] & land area in square miles \\
67 & 0.3515 & [0.3106, 0.4052] & population density in persons per square mile \\ 
9  & 0.1004 & [0.0589, 0.1498] & \# of people living in areas classified as urban \\ 
33 & 0.0751 & [0.0412, 0.1058] & \% of kids in family housing with two parents  \\ 
47 & 0.0272 & [0.0140, 0.0417] & \% of persons in dense housing (more than 1 person per room) \\ 
32 & 0.0242 & [0.0012, 0.0581] & \% of families (with kids) that are headed by two parents \\ 
13 & 0.0242 & [0.0094, 0.0451] & \% of households with investment / rent income in 1989 \\ 
4  & 0.0163 & [0.0029, 0.0329] & \% of population that is caucasian  \\ 
1  & 0.0153 & [0.0003, 0.0394] & population for community  \\ 
3  & 0.0137 & [0.0014, 0.0278] & \% of population that is african american \\ 
15 & 0.0080 & [0.00002, 0.0165] & \% of households with public assistance income in 1989  \\ 
6  & 0.0080 & [0.0004, 0.0195] & \% of population that is of hispanic heritage  \\ 
43 & 0.0053 & [0.0013, 0.0125] & \% of people who do not speak English well \\ 
68 & 0.0036 & [0.0004, 0.0072] & \% of people using public transit for commuting \\ 
49 & 0.0031 & [0.0001, 0.0101] & \# of vacant households  \\ 
50 & 0.0031 & [0.0007, 0.0067] & \% of housing occupied \\ 
62 & 0.0027 & [0.0002, 0.0068] & \# of people in homeless shelters \\ 
38 & 0.0025 & [0.00007, 0.0103] & \# of kids born to never married  \\ 
45 & 0.0024 & [0.0004, 0.0072] & \% of all occupied households that are large (6 or more people) \\ 
44 & 0.0023 & [0.0005, 0.0068] & \% of family households that are large (6 or more) \\ 
31 & 0.0020 & [0.00009, 0.0077] & mean number of people per family \\ 
41 & 0.0018 & [0.00009, 0.0051] & \% of population who have immigrated within the last 5 years \\ 
5  & 0.0017 & [0.00008, 0.0042] & \% of population that is of asian heritage  \\ 
23 & 0.0013 & [0.00008, 0.0038] & \% of people 16 and over, in the labor force, and unemployed \\ 
\hline
\end{tabular}}
\begin{tabnote}
$j$, $j$-th predictor; Mean, posterior mean; 90\%CI refers to a 90\% credible interval.
\end{tabnote}
\label{tb:list11}
\end{table}
%%%%%%%%%%%%%%%%%%%%%%%%%%%%%%%%%%%%%%%%%%%%%%%%%%%%%%%%%%%%%%%%%%%%%%%%%%%%%%%%%%%%%%%%%%%%

%%%%%%%%%%%%%%%%%%%%%%%%%%%%%%%%%
\begin{table}[htp]
\tbl{List of the selected predictors by the proposed method  in descending order of the posterior means of conditional mutual information with non-violent crimes as the response}{%
\begin{tabular}{cccl}
\hline
\multicolumn{1}{c}{$j$} & \multicolumn{1}{c}{Mean}	& \multicolumn{1}{c}{90\%CI}	& \multicolumn{1}{l}{Predictor}	\\
\hline 
66 & 0.9859 & [0.9500, 1.0189] & land area in square miles \\
67 & 0.8282 & [0.7870, 0.8700] & population density in persons per square mile \\ 
32 & 0.0300 & [0.0082, 0.0518] & \% of families (with kids) that are headed by two parents \\ 
28 & 0.0217 & [0.0011, 0.0475] & \% of males who have never married \\ 
33 & 0.0217 & [0.0015, 0.0484] & \% of kids in family housing with two parents  \\ 
9  & 0.0200 & [0.0017, 0.0518] & \# of people living in areas classified as urban \\ 
30 & 0.0183 & [0.0006, 0.0399] & \% of population who are divorced \\ 
27 & 0.0182 & [0.0001, 0.0443] & \% of males who are divorced \\ 
47 & 0.0181 & [0.0043, 0.0353] & \% of persons in dense housing (more than 1 person per room) \\ 
1  & 0.0174 & [0.0001, 0.0426] & population for community  \\ 
29 & 0.0086 & [0.0003, 0.0223] & \% of females who are divorced  \\ 
64 & 0.0072 & [0.0007, 0.0155] & \% of people born in the same state as currently living  \\ 
50 & 0.0039 & [0.0010, 0.0075] & \% of housing occupied \\ 
52 & 0.0023 & [0.0001, 0.0058] & \% of vacant housing that is boarded up \\ 
\hline
\end{tabular}}
\begin{tabnote}
$j$, $j$-th predictor; Mean, posterior mean; 90\%CI refers to a 90\% credible interval.
\end{tabnote}
\label{tb:list12}
\end{table}
%%%%%%%%%%%%%%%%%%%%%%%%%%%%%%%%%%%%%%%%%%%%%%%%%%%%%%%%%%%%%%%%%%%%%%%%%%%%%%%%%%%%%%%%%%%%

%%%%%%%%%%%%%%%%%%%%%%%%%%%%%%%%%
%\section{90\% credible intervals of CMIs in the criminology application}
%%%%%%%%%%%%%%%%%%%%%%%%%%%%%%%%%

%90\% credible intervals of CMIs are reported in Figure \ref{fig1} ( assaults, burglaries and larcenies), Figure \ref{fig2} (auto thefts, arsons and violent crimes) and Figure \ref{fig3} (non-violent crimes). 

\clearpage

\begin{figure}[htbp]
\figurebox{18pc}{40pc}{}[CMI-rape2.eps]
\caption{90\% credible intervals of the estimated conditional mutual information with rapes as the response for each of the 68 demographic predictors adjusting for the other predictors.}
\label{fig1}
\end{figure}

\begin{figure}[htbp]
\figurebox{18pc}{40pc}{}[CMI-robbery2.eps]
\caption{90\% credible intervals of the estimated conditional mutual information with robberies as the response for each of the 68 demographic predictors adjusting for the other predictors.}
\label{fig2}
\end{figure}

\begin{figure}[htbp]
\figurebox{18pc}{40pc}{}[CMI-assault2.eps]
\caption{90\% credible intervals of the estimated conditional mutual information with assaults as the response for each of the 68 demographic predictors adjusting for the other predictors.}
\label{fig3}
\end{figure}

\begin{figure}[htbp]
\figurebox{18pc}{40pc}{}[CMI-burglary2.eps]
\caption{90\% credible intervals of the estimated conditional mutual information with burglaries as the response for each of the 68 demographic predictors adjusting for the other predictors.}
\label{fig4}
\end{figure}

\begin{figure}[htbp]
\figurebox{18pc}{40pc}{}[CMI-larceny2.eps]
\caption{90\% credible intervals of the estimated conditional mutual information with larcenies as the response for each of the 68 demographic predictors adjusting for the other predictors.}
\label{fig5}
\end{figure}

\begin{figure}[htbp]
\figurebox{18pc}{40pc}{}[CMI-autotheft2.eps]
\caption{90\% credible intervals of the estimated conditional mutual information with auto thefts as the response for each of the 68 demographic predictors adjusting for the other predictors.}
\label{fig6}
\end{figure}

\begin{figure}[htbp]
\figurebox{18pc}{40pc}{}[CMI-arson2.eps]
\caption{90\% credible intervals of the estimated conditional mutual information with arsons as the response for each of the 68 demographic predictors adjusting for the other predictors.}
\label{fig7}
\end{figure}

\begin{figure}[htbp]
\figurebox{18pc}{40pc}{}[CMI-violent2.eps]
\caption{90\% credible intervals of the estimated conditional mutual information with violent crimes as the response for each of the 68 demographic predictors adjusting for the other predictors.}
\label{fig8}
\end{figure}

\begin{figure}[htbp]
\figurebox{18pc}{40pc}{}[CMI-nonviolent2.eps]
\caption{90\% credible intervals of the estimated conditional mutual information with non-violent crimes as the response for each of the 68 demographic predictors adjusting for the other predictors.}
\label{fig9}
\end{figure}

\clearpage

%%%%%%%%%%%%%%%%%%%%%%%%%%%%%%%%%
%\section{Lists of the selected predictors for competitors in the criminology application}
%%%%%%%%%%%%%%%%%%%%%%%%%%%%%%%%%

%%%%%%%%%%%%%%%%%%%%%%%%%%%%%%%%%
\begin{table}[htp]
\tbl{List of the selected predictors for murders, rapes, robberies, assaults, burglaries and larcenies by the competitors}{%
\begin{tabular}{cl}
\hline
\multicolumn{2}{l}{Murder:}	\\
\multicolumn{1}{c}{Method} & \multicolumn{1}{l}{Variable numbers of the selected predictors}	\\
\hline 
LASSO & 19, 38, 39, 49 \\ 
CM & all variables  \\ 
NCCO & 1,3,4,5,6,7,8,9,11,13,14,15,19,20,21,22,23,24,25,26,27,28,29,30,32,33, \\
& 34,35,36,37,38,39,40,41,42,43,44,45,46,47,48,49,51,52,61,64,65,68 \\ 
AQM & 3, 4, 13, 38, 49, 53, 64 \\ 
\hline
\multicolumn{2}{l}{Rape:}	\\
\multicolumn{1}{c}{Method} & \multicolumn{1}{l}{Variable numbers of the selected predictors}	\\
\hline 
LASSO & 1, 3, 9, 16, 27, 28, 32, 35, 38, 49, 50, 52, 54, 66, 67 \\ 
CM & all variables  \\ 
NCCO & 1,3,4,5,6,7,8,9,11,13,14,15,19,20,21,22,23,24,25,26,27,28,29,30,32,33,
\\
& 34,35,36,37,38,39,40,41,42,43,44,45,46,47,48,49,51,52,61,64,65,68 \\ AQM & 3, 4, 13, 38, 49, 53, 64 \\
\hline
\multicolumn{2}{l}{Robbery:}	\\
\multicolumn{1}{c}{Method} & \multicolumn{1}{l}{Variable numbers of the selected predictors}	\\
\hline 
LASSO & 2, 4, 15, 25, 28, 31, 38, 39, 41, 44, 49, 50, 52, 62, 63 \\ 
CM & all variables  \\ 
NCCO & 1,3,4,5,6,7,8,9,11,13,14,15,19,20,21,22,23,24,25,26,27,28,29,30,32,33,
\\
& 34,35,36,37,38,39,40,41,42,43,44,45,46,47,48,49,51,52,61,64,65,68 \\
AQM & 3, 4, 13, 32, 33, 34, 35, 36, 38, 39, 46, 48, 51, 53, 64
 \\ 
\hline
\multicolumn{2}{l}{Assault:}	\\
\multicolumn{1}{c}{Method} & \multicolumn{1}{l}{Variable numbers of the selected predictors}	\\
\hline 
LASSO & 1, 38, 39 \\
CM & all variables  \\ 
NCCO & 1,3,4,5,6,7,8,9,11,13,14,15,19,20,21,22,23,24,25,26,27,28,29,30,32,33,
\\
& 34,35,36,37,38,39,40,42,43,44,45,46,47,48,49,51,52,61,64,65,68 \\
AQM & 3, 4, 13, 22, 26, 38, 49 \\ 
\hline
\multicolumn{2}{l}{Burglary:}	\\
\multicolumn{1}{c}{Method} & \multicolumn{1}{l}{Variable numbers of the selected predictors}	\\
\hline 
LASSO & 1, 3, 4, 9, 16, 19, 25, 27, 33, 43, 49, 50, 52, 53, 64, 66, 67  \\ 
CM & all variables  \\ 
NCCO &  1,3,4,5,6,7,8,9,11,13,14,15,19,20,21,22,23,24,25,26,27,28,29,30,32,33,
\\
& 34,35,36,37,38,39,40,41,42,43,44,45,46,47,48,49,51,52,61,64,65,68 \\
AQM & 4, 13, 19, 22, 38, 49, 64 \\ 
\hline
\multicolumn{2}{l}{Larceny:}	\\
\multicolumn{1}{c}{Method} & \multicolumn{1}{l}{Variable numbers of the selected predictors}	\\
\hline 
LASSO & 1, 6, 9, 16, 19, 22, 25, 27, 28, 30, 49, 50, 53, 60, 66, 67 \\ 
CM & all variables  \\ 
NCCO &   1,3,4,5,6,7,8,9,11,13,14,15,19,20,21,22,23,24,25,26,27,28,29,30,32,33,
\\
& 34,35,36,37,38,39,40,42,43,44,45,46,47,48,49,51,52,53,61,64,65,68 \\
AQM & 19, 36, 37, 40, 46, 48, 51, 53, 64, 65 \\ 
\hline
\end{tabular}}
\begin{tabnote}
CM, Cram\'{e}r-von-Mises type statistic; NCCO, normalized cross-covariance operator; AQM, asymmetric quadratic measure.
\end{tabnote}
\label{tb:list13}
\end{table}
%%%%%%%%%%%%%%%%%%%%%%%%%%%%%%%%%%%%%%%%%%%%%%%%%%%%%%%%%%%%%%%%%%%%%%%%%%%%%%%%%%%%%%%%%%%%

%%%%%%%%%%%%%%%%%%%%%%%%%%%%%%%%%
\begin{table}[htp]
\tbl{List of the selected predictors for auto thefts, arsons, violent crimes and non-violent crimes by the competitors}{%
\begin{tabular}{cl}
\hline
\multicolumn{2}{l}{Auto Thef:}	\\
\multicolumn{1}{c}{Method} & \multicolumn{1}{l}{Variable numbers of the selected predictors}	\\
\hline 
LASSO & 1, 9, 19, 39 \\
CM & all variables  \\ 
NCCO &  1,3,4,5,6,7,8,9,11,13,14,15,19,20,21,22,23,24,25,26,27,28,29,30,32,33,
\\
& 34,35,36,37,38,39,41,42,43,44,45,46,47,48,49,51,52,61,64,65,68 \\
AQM & 3, 4, 13, 22, 26, 38, 39, 48, 51, 53 \\ 
\hline
\multicolumn{2}{l}{Arson:}	\\
\multicolumn{1}{c}{Method} & \multicolumn{1}{l}{Variable numbers of the selected predictors}	\\
\hline 
LASSO & 1, 19, 39 \\ 
CM & all variables  \\ 
NCCO &   1,3,4,5,6,7,8,9,11,13,14,15,19,20,21,22,23,24,25,26,27,28,29,30,32,33,
\\
& 34,35,36,37,38,39,40,41,42,43,44,45,46,47,48,49,51,52,61,64,65,68 \\
AQM & 53 \\ 
\hline
\multicolumn{2}{l}{Violent Crime:}	\\
\multicolumn{1}{c}{Method} & \multicolumn{1}{l}{Variable numbers of the selected predictors}	\\
\hline 
LASSO & 3, 5, 9, 25, 27, 32, 35, 39, 41, 42, 52, 62, 63, 66, 67 \\ 
CM & all variables  \\ 
NCCO &    1,3,4,5,6,7,8,9,11,13,14,15,19,20,21,22,23,24,25,26,27,28,29,30,32,33,
\\
& 34,35,36,37,38,39,40,42,43,44,45,46,47,48,49,51,52,61,64,65,68 \\
AQM & 3, 4, 13, 22, 26, 36, 38, 53, 64 \\ 
\hline
\multicolumn{2}{l}{Non-Violent Crime:}	\\
\multicolumn{1}{c}{Method} & \multicolumn{1}{l}{Variable numbers of the selected predictors}	\\
\hline 
LASSO & 1, 9, 19, 30, 49, 66  \\ 
CM & all variables  \\ 
NCCO &     1,3,4,5,6,7,8,9,11,13,14,15,19,20,21,22,23,24,25,26,27,28,29,30,32,33,
\\
& 34,35,36,37,38,39,42,43,44,45,46,47,48,49,51,52,53,61,64,65,68 \\
AQM & 19, 36, 51, 53 \\ 
\hline
\end{tabular}}
\begin{tabnote}
CM, Cram\'{e}r-von-Mises type statistic; NCCO, normalized cross-covariance operator; AQM, asymmetric quadratic measure.
\end{tabnote}
\label{tb:list14}
\end{table}
%%%%%%%%%%%%%%%%%%%%%%%%%%%%%%%%%%%%%%%%%%%%%%%%%%%%%%%%%%%%%%%%%%%%%%%%%%%%%%%%%%%%%%%%%%%%

\bibliographystyle{biometrika}
\bibliography{cmi2}